\documentclass[useAMS,usenatbib,asm]{mn2e}
\usepackage{graphicx,fleqn,times,color}
\newcommand{\tal}{\it et al. \rm}
\newcommand{\ksp}{km\,s^{-1}\,kpc^{-1}}
\newcommand{\gr}{^{\circ}}
\title{Modelling the inner disc of the Milky Way with manifolds. I - A 
first step.}
\author[Romero-G\'omez \tal]{M. Romero-G\'omez$^1$, E. Athanassoula$^2$, 
T. Antoja$^3$, F. Figueras$^1$  \\  \\
$^1$Dept. d'Astronomia i Meteorologia, Institut de Ci\`encies del Cosmos (ICC), Universitat de Barcelona (IEEC-UB), Mart\'{i} i Franqu\`es 1, E08028 Barcelona, Spain\\
$^2$Laboratoire d'Astrophysique de Marseille (LAM), UMR6110, 
CNRS/Universit\'e de Provence,\\
Technop\^ole de Marseille Etoile, 38 rue Fr\'ed\'eric Joliot Curie, 13388 Marseille C\'edex 20, France\\
$^3$Kapteyn Astronomical Institute, University of Groningen, PO Box 800, 
9700 AV Groningen, the Netherlands\\
}
\date{Received }

\begin{document}

\maketitle

\begin{abstract}
We study the bar-driven dynamics in the inner 
part of the Milky Way by using invariant manifolds. This theory has
been successfully applied to describe the morphology and kinematics of 
rings and spirals in external galaxies, and now, for the first time,
we apply it to the 
Milky Way. In particular, we compute the orbits confined by the
invariant manifolds of the unstable periodic orbits located at the
ends of the bar. We start by discussing whether the COBE/DIRBE bar and
the Long bar compose a single bar or two independent bars
and perform a number of comparisons which, taken together, argue
strongly in 
favour of the former. More specifically, we favour the possibility
that the so-called COBE/DIRBE bar is the boxy/peanut bulge of a bar
whose outer thin parts are the so-called Long bar. This possibility is
in good agreement both with observations of external galaxies, with
orbital structure theory and with simulations. We then analyse 
in detail the morphology and kinematics given by five representative Galactic 
potentials. Two of these have a Ferrers bar, two have a quadrupole bar
and the last one a composite bar. We first consider only the COBE/DIRBE
bar and then extend it to include the effect of the Long bar. We find that 
The large-scale structure given by the manifolds describes 
an inner ring, whose size is similar to the near and far 3-kpc arm, and an 
outer ring, whose properties resemble those of the Galactic Molecular Ring. 
We also analyse the kinematics of these two structures, under the
different galactic potentials, and find they reproduce the relevant
over-densities found in the  
galactic longitude - velocity CO diagram. Finally, we consider for what 
model parameters, the global morphology of the manifolds may reproduce the two 
outer spiral arms. We conclude that this would necessitate either more
massive and more rapidly rotating bars, or including in the potential an
extra component describing the spiral arms. 
\end{abstract}


\section{Introduction}
\label{sec:intro}
The large-scale structure of the Milky Way (hereafter, MW) disc has been under 
study for many years. The COBE/DIRBE \citep{wei94} and Spitzer/GLIMPSE 
\citep{chu09} missions provided infrared information on the global structure 
of the inner Galaxy. Even though these studies have provided some light, the 
large-scale structure of the MW disc proves to be highly complex.  
Near- and mid-IR low resolution images detected the COBE/DIRBE bar 
\citep{wei94}, also referred to as the triaxial bulge or the COBE/DIRBE bar. 
Near-IR red clump giants of the mid plane revealed the existence of a second 
bar \citep{ham00,lop07,cab08}, and confirmed by GLIMPSE \citep{ben05}, usually
referred to as the Long bar. Observations suggest two other large scale 
structures towards the inner parts of the Milky Way, namely the near and far 
3-kpc arms \citep{ker64,dam08} and the Galactic Molecular Ring \citep{cle88}.
Although their characteristics or even their existence are currently being
under debate \citep{dam11}, here we aim to bring some light given the observed
characteristics up-to-date. The near and far 3-kpc arms were
detected using the HI 21-cm line and CO emission surveys  
and extend roughly parallel to the COBE/DIRBE bar, whereas the position of the
Galactic Molecular Ring (hereafter, GMR) is not so well determined. 
\citet{cle88} suggested it is located at $\sim 5.5\,kpc$ from the Galactic 
Center, while other authors suggest it is located about halfway to the 
Galactic Center \citep{bin98,dam01,rat09}.

There has been a lot of effort to determine the bar and spiral arm
characteristics, both from the observational and theoretical point of view. 
\citet{deb02} and \citet{sev02} use a sample of $\sim 250$ bright
OH/IR stars of the inner Galaxy to determine the pattern speed of the 
COBE/DIRBE bar, while the CO emission and the 21-cm line of neutral 
hydrogen map the galactic longitude-velocity (l,v) diagram 
\citep{dam01,dam08,val08}. The spiral arms and rings of the Galaxy appear as 
over-densities in such diagrams. From the theoretical point of view,
several works use hydrodynamics simulations to constrain the Galaxy parameters 
and to reproduce the observed (l,v) diagram
\citep[e.g.][]{eng99,rod08,bab10}. From the orbits point of view, \citet{hab06}
compute a library of orbits, some of which reproduce the over-densities in the
inner longitudes of the ($l$,$v$)-diagram, while \citet{gre11} study in detail 
the shape of the 3-kpc arm concluding that it can be approximated by an 
elliptical ring. Test particle simulations have recently been used to study 
the velocity distribution function in the Solar neighbourhood and to use it 
to constrain the characteristics of the bar and/or spiral arms 
\citep{deh00,fux01,cha07,ant09,min10}. 

In this paper we will use an approach which is novel for our Galaxy,
namely that of invariant manifolds. In a previous set of papers we
developed the method and techniques that we will use here and also
applied them to the study of spirals and rings in external galaxies
\citep[see][hereafter Papers I-V, respectively]{rom06, rom07, 
ath09, ath09b, ath10}. The invariant manifolds are linked to the presence of
the Lagrangian points $L_1$ and $L_2$ of a barred system and they 
can reproduce the observed structures of rings and spiral arms. 
In papers I-V, we studied in detail the characteristics of the orbits 
confined by such invariant manifolds and we analysed both their
morphology and kinematics so as to compare them to the rings and
spiral arms in external barred  galaxies. Invariant manifolds, albeit 
in a quite different way than what we have here and Papers I-V, have also been 
used by \citet{tso09} to model three barred galaxies, NGC~3992, NGC~1073, and 
NGC~1398 and by \citet{PatsisKG2010} to perform an orbital analysis of 
NGC~1300.

Here we use the invariant manifolds to compute a 
family of orbits of a wide range of energies and to study their morphology 
and their kinematics. We want to evaluate whether the invariant manifolds 
can provide an alternative, plausible model for the inner part of the MW disc. 
By using MW analytical potentials recently used in the literature, we try to 
answer the following questions: Can observations be plausibly
interpreted by manifolds? Can the latter provide an alternative
interpretation of the inner structure 
of the MW? Which are the requirements the potential has to fulfill in order
to reproduce the rings and spiral arms of the MW? In a future paper we
will examine 
whether the combination of the manifolds and observations can constrain the 
MW bar properties. 

In order to compute the orbits confined by the manifolds we must 
first fix the Galactic potential. We have chosen the most representative 
analytical potentials used to describe a COBE/DIRBE bar, namely a Ferrers 
bar \citep{fer77}, a quadrupole bar \citep{bin08} and a composite bar 
\citep{pic04}. Deliberately, we consider studies where the authors 
have tuned one of these three types of potentials to the COBE/DIRBE bar,
and we use the same set of parameters as they do.

The paper is organised as follows: in Sect.~\ref{sec:modelpot}, we
discuss whether our Galaxy has a single or a double bar, using
arguments from the morphology of external galaxies, from orbital structure 
theory and from $N$-body simulations. We also describe 
the models and compare them in detail in terms of forces. In 
Sect.~\ref{sec:motion}, we give a brief summary of the dynamics driven by 
the unstable Lagrangian points and, in particular, the definition of the 
invariant manifolds. We also give a brief summary of the main relevant 
results found in papers I-V. In Sect.~\ref{sec:inner}, we compute the 
invariant manifolds for the selected models and we analyse them in terms of 
morphology and kinematics. The results are compared to the observables in 
Sect.~\ref{sec:MW}. In Sect.~\ref{sec:outer}, we explore the parameter space 
and determine in which cases the manifolds could reproduce outer spiral arms. 
Finally, we give a short summary and conclusions in Sect.~\ref{sec:conc}. 
In the Appendix, we describe in detail the analytical models and give the 
default parameters used.

\section{Modelling the Galactic potential}
\label{sec:modelpot}

\subsection{Analytical models}
\label{sec:analyticmod}

There are several analytical models in the literature used to model the MW 
Galaxy. They essentially consist of an axisymmetric plus a one-bar component. 
Each model has been constructed to model the Milky Way and in the Appendix of 
this paper we give a brief description of the potentials used and their default 
parameters. We want to stress here that we will consider the same parameters 
as these studies. The axisymmetric component describes the disc, 
halo and bulge of the Galaxy and in each model it is modelled in a different
way.

The models considered in this paper are:
\begin{itemize}
\item Melnik \& Rautiainen (2009, hereafter MR09) and Gardner \& Flynn (2010, 
hereafter GF10). Both use a Ferrers bar \citep{fer77}, though the purpose of 
each of the papers is very different. The former uses the bar potential in test 
particle simulations to model the kinematics of the outer rings and spirals 
of the Galaxy and to compare it with the residual velocities of 
OB-associations in the Perseus and Sagittarius regions. The latter studies 
the effect of the bar parameters on the kinematic substructures found in the 
velocity plane of the Solar neighbourhood. 
\item \citet{deh00} and \citet{fux01} (hereafter, Dehnen00 and Fux01, 
respectively). Both use a quadrupole bar, but with different model parameters,
to study the effect of the COBE/DIRBE bar on the local disc stellar kinematics.
Such potentials are also often referred to as ad-hoc potentials, 
since they are not obtained from any density distribution.
\item The composite bar of Pichardo {\it et al.} (2004, hereafter PMM04) 
consists of a set of prolate ellipsoids, superposed so that the surface 
density matches the mass distribution obtained by the COBE/DIRBE mission, 
and from which the potential and forces are derived. 
The authors compute in detail the families of orbits given by this potential 
and use surfaces of section to characterise the bar structure.
\end{itemize}

In order to fix the parameters of each model, the authors take into
account the available relevant observational data.
Even though there is a lot of uncertainty in these data, a range of possible 
values can be determined.  
The semi-major axis of the COBE/DIRBE bar or Galactic bulge is $\sim
3.1-3.5\,kpc$ and its aspect ratio of  
$10:4:3$ (length:width:height), as estimated by COBE/DIRBE \citep{wei94,fre98,
ger02}. Several studies fix the mass of the COBE/DIRBE bar in 
the range $1-2\times 10^{10}\,M_{\odot}$ 
\citep{mat82,ken92,dwe95,zha96,wei99}. The relative orientation of
the bar with 
respect to the Galactic Center - Sun is not well established, although most 
observations (2MASS star counts or red clump giants) and models (based 
on COBE/DIRBE and Spitzer/GLIMPSE) agree it lies roughly within the range of 
$15\gr-30\gr$ (e.g. 
\citet{dwe95,bin97,sta97,fux99,eng99,bis02,bis03,bab05,lop05,ben05,chu09}).
Here we choose $20\gr$ as a representative value. The pattern speed obtained 
by the studies mentioned in the Introduction lies within the range 
$\Omega_b=35-60\ksp$ although higher valuers are favored (see \citealt{ger10} 
for a review). As for the Long bar, 
it is somewhat longer than the COBE/DIRBE bar with a semi-major axis of 
$\sim 4-4.5\,kpc$ and an aspect ratio of $10:1.54:0.26$ \citep{ham00}. The mass 
of the Long bar is less than the mass of the COBE/DIRBE bar. It is estimated 
to be around $6\times 10^9\,M_{\odot}$, i.e. about $2/3$ that of the COBE/DIRBE 
bar \citep{ham00,gar10}. The relative orientation from the Galactic Center - 
Sun line is estimated observationally to be $\sim 40\gr$ 
\citep{ham00,ben05,lop07,cab08}.
Note that the observational sources used to constrain the parameters of
the COBE/DIRBE and the Long bar are different. These could, in principle, 
lead to somewhat different definitions of bar length. However, according to 
\citet{fre98} for the COBE/DIRBE bar and \citet{ham00} for the Long bar, both 
give estimates of the bar half length. Even though these estimates are subject
to observational errors, the ratio of the two values will not change 
significantly.

\subsection{How many bars does our Galaxy have?}
\label{sec:1or2bars}

A considerable fraction of external barred galaxies are known to have
two bars: a primary, or main bar and a secondary, or inner bar. This
fraction depends on the galaxy's Hubble type, the quality 
of the sample images, and other factors, but fractions of the order 
of a third or a fourth are quite reasonable. Could it be 
that our Galaxy is one of these? In order to pursue
this line further, one needs to make sure that the properties of the
COBE/DIRBE bar and the Long bar are compatible with those of galaxies 
with double bars. Can we safely assume that the Long bar is the main bar and 
that the COBE/DIRBE bar is the secondary bar? 

Several properties of inner bars have been
well studied (e.g. \citealt{Erwin.Sparke.02}, \citealt{Laine.SKP02},
  \citealt{Erwin.09}) and some major trends have been found. The
strongest constraints come from the bar length. Inner
bars are quite small, with a semi-major axis between $100$ pc and $1.2\,kpc$,
with median size around $500\,pc$. Typically their relative length is
about 12\% of that of the main bar. The sample of \cite{Erwin.09} 
contains 64 galaxies with double bars and out of these only two have a
secondary bar longer than 22\% of the primary, and none
longer than 30\%. This contrasts strongly with the
numbers for our Galaxy, where the length of the bar semi-major axis are
3.1 -- 3.5 $kpc$ and $\sim 4\,kpc$ for the COBE/DIRBE and for the Long
bar, respectively. Thus the length of the COBE/DIRBE bar is
more than 10 sigmas beyond the values found for the inner bars of external
galaxies. Furthermore, the relative length of the COBE/DIRBE bar
relative to the Long bar is $\sim 0.8$, again more than 10 
sigmas out of the distribution found from external galaxies. 
Double bars have also been found in simulations \citep{Heller.SA07,
  Shen.Debattista.09} and their parameters are in
good agreements with those of observed double bars and in disagreement
with the values for our Galaxy. 

So one can reasonably exclude that the Long and the COBE/DIRBE bar
form a double bar system, because their properties are very far from
those of double bars in external galaxies and simulations, making the
two incompatible\footnote{\citet{Alard01}, however, found evidence
  for a small lopsided bar in our Galaxy, whose size is well
  compatible with those of inner bars. This would then be the
  secondary bar of the Galactic double bar system, while the primary
  would be constituted of the 
  COBE/DIRBE bar and the Long bar together.}. So then what is it? 
The data give us an important 
clue for that. Namely the ratio of the major- to z- semi-axis of the bar is
$\sim 0.3$ for the COBE/DIRBE bar and $\sim 0.026$ for the Long bar,
i.e. {\it the Long bar is very thin} and {\it the COBE/DIRBE bar is very
thick}. Let us therefore examine the 
alternative that there is only one bar in the galaxy and that the COBE/DIRBE
bar is simply the part which corresponds to the boxy/peanut bulge and
the Long bar the outer part of this bar. This geometry has been already 
discussed for external galaxies in \citep[e.g.][]{Athanassoula05,
Athanassoula.Beaton.06} and was first proposed for our Galaxy by
\cite{Athanassoula06,Athanassoula08}. \cite{cab07} tested this suggestion using
their red-clump giants measurements. We will discuss here further what
the relevant orbital structure studies and $N$-body simulations imply. 

\begin{figure}
\centering
\includegraphics[scale=0.15,angle=0.]{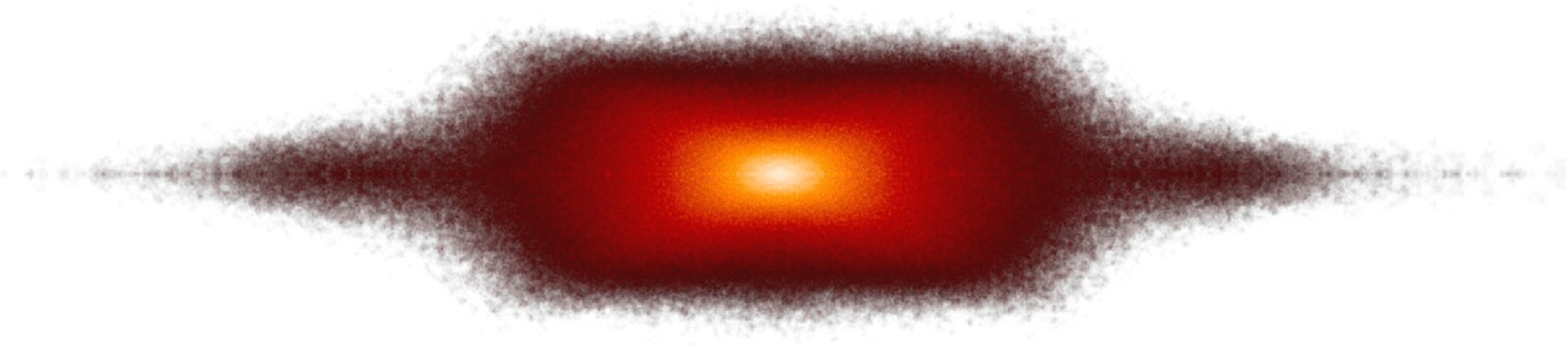}
\hspace{0.2cm}
\includegraphics[scale=0.15,angle=0.]{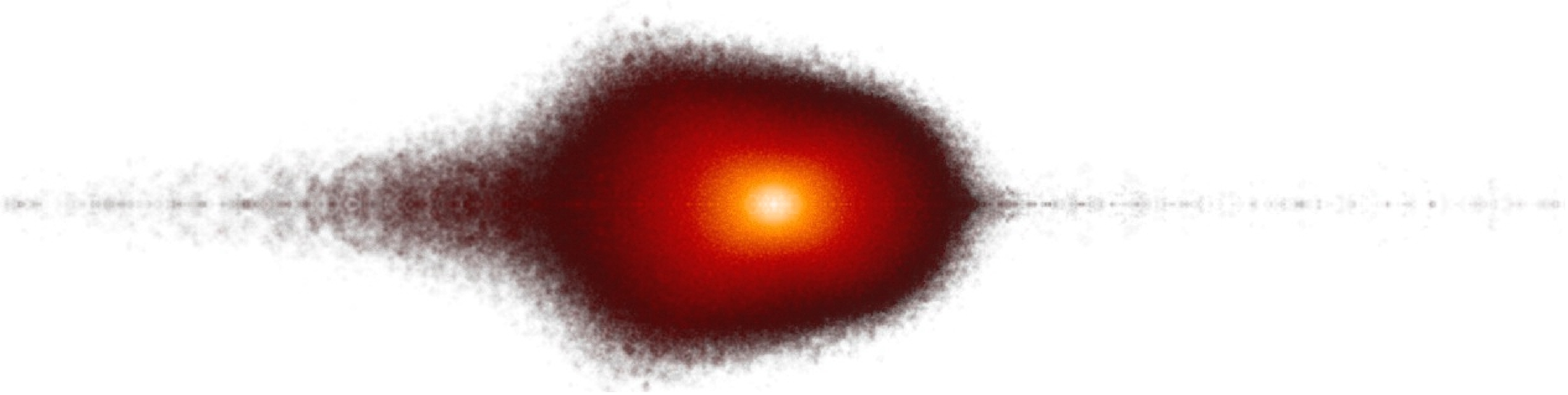}
\hspace{0.15cm}
\caption{Two edge-on views of a bar from a simulation (see text). The
  upper panel shows a side-on view, and the bottom one a view from an
  angle near the bar major axis. In both cases the views are
  perspective and thus the relative thickness of the inner and outer 
parts should be inferred only from the upper panel.  
}
\label{fig:1or2bars}
\end{figure}

\begin{figure}
\centering
\includegraphics[scale=0.3,angle=0.]{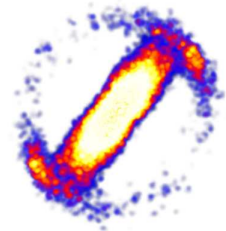}
\hspace{0.15cm}
\caption{Face-on view of a simulation. This is not a model built specifically
to represent our Galaxy, but is a clear example of a snapshot with a
short leading ring segment emanating from the end of the
bar (see text). In this figure the rotation is clockwise to facilitate
comparison with our Galaxy.
}
\label{fig:leading}
\end{figure}

\citet{Pfenniger84} and \citet{sko02a,sko02b} studied the building
blocks of bars, i.e. the periodic orbits, in 3D. They found that the
third dimension introduced considerable complexity to the orbital
structure. Whereas in 2D it is the orbits of the $x_1$ family that are the
backbone of the bar \citep{Contopoulos.Papayannopoulos.80, 
ath83}, in 3D we have a tree of 2D and 3D families
bifurcating from the $x_1$ \citep{sko02b}. Each of these
families has its own horizontal and vertical extent. Since the extent
of the box/peanut will in general be determined by a {\it different family}
than that determining the length of the bar, it is natural for their
lengths to be different. Furthermore, the ratio of the lengths
predicted in this way \citep{PatsisSA02}, is in good agreement with
that measured for the COBE/DIRBE bar and the Long bar (0.8). 

This structure has also been seen in a number of simulations, where the bar 
forms very thin and after a while a vertical instability develops and
creates the boxy/peanut feature \citep[e.g.][]{Binney81, Combes.DFP.90,
  Athanassoula05, Martinez.VSH.06}. Results from one such simulation
are shown in Fig.~\ref{fig:1or2bars} (see Appendix for the details of the 
simulation). This is given as an illustration and not as a model of our
Galaxy. In the upper panel we give the side-on 
view\footnote{In a side-on view the galaxy is
  viewed edge-on, with the line of
sight perpendicular to the bar major axis} of the bar component. For
clarity, the remaining disc, as well as the halo are not displayed.
This shows clearly that the thick part of the bar (i.e. the
boxy/peanut bulge)
is less extended than the thin part and that the thin part protrudes
on either side of it. It also gives an estimate of the relative vertical 
thickness of the inner and outer parts, although this could vary from one
model to another. Note that the thin and thick part are {\it parts
  of the same bar}, and do not constitute two separate
components. Furthermore, both orbital theory 
and simulations show that, in the present scenario of box/peanut
formation, there {\it must} be a thin part of the bar extending further
than the box/peanut. 

In the lower panel of Fig.~\ref{fig:1or2bars} the bar is viewed from
an angle much nearer to  
end-on\footnote{In the end-on view the galaxy is viewed edge-on with the
line of sight along the bar major axis}. We do not claim that either
the model, or the angle of the
line of sight with respect to the bar major axis is necessarily the
correct one. This projection, nevertheless, illustrates roughly how one could be
mistaken into considering the poxy/peanut feature and the thin outer
part of this bar as two separate components\footnote{A short movie, showing
this bar from several viewing angles can be found in 
\texttt{lam.oamp.fr/recherche-14/dynamique-des-galaxies/}
\texttt{scientific-results/peanuts/milkyway/movies-98/}. The username and
the password are RefereeArticle and peanut, respectively.}.   

By visualising a simulation from many viewing angles, it is possible to
realise the geometry of the object, but this is not possible
for real galaxies, where only one viewing angle is
possible for each case. Thus, in near face-on galaxies we can clearly
see the bar, while in near side-on ones we can see the boxy/peanut
bulge. There are, however, viewing angles which are
near edge-on, but not quite, and where both the boxy/peanut shape and the
outer thin bar are visible. The most interesting inclination range is
between $60^{\circ}$ and $80^{\circ}$. Several examples have already been
discussed in the literature, such as NGC 7582 with an inclination
angle of about $65^{\circ}$ \citep{Quillen.KFD97}, NGC 4442
\citep{Bettoni.Galletta.94} at approximately $72^{\circ}$, and M31
at about $77^{\circ}$ \citep{Athanassoula.Beaton.06}. In these cases
it is possible to get information on the ratio of lengths of the
box/peanut and bar. For M31, where this was specifically measured with
the help of cuts \citep{Athanassoula.Beaton.06}, it was found to be
$\sim 0.7$, in good agreement with the value found for our
Galaxy. \cite{Lutticke.DP00} made a detailed morphological and
photometrical study of a sample of 60 edge-on, also using cuts, and
found that the ratio of boxy/peanut length to bar length depends on
the specific morphology of boxy/peanut bulge and, therefore, on bar
strength. For peanuts, this ratio is $0.53 \pm 0.08$, for clear boxes
$0.63 \pm 0.08$ and for box-like shapes $0.71 \pm 0.1$. According to these 
numbers our Galaxy would be more box-like, but it should be kept in mind 
that these statistics are based on very few objects (21 in total, for this 
measurement) and the scatter quite high. 

The above arguments seem to exclude the possibility that the Long bar
and the COBE/DIRBE bar are the primary and secondary bars of
a double bar system, since the
bar lengths and length ratios disagree strongly both with observations of
external galaxies and with simulations. Double systems with bars of
comparable length have never been observed either in any external
galaxy or in any simulation. It would thus be very hazardous to
assume that our Galaxy is the only one known to have such a
feature. On the other hand, the 
alternative that the COBE/DIRBE and the Long bar are parts of a single
bar is in good agreement with observations, with orbital structure and
with simulations. It is thus reasonable to favour this second
alternative. 

Yet one inconsistency could still remain, concerning the position
angles of the bar(s). The angle between the major axis of the
COBE/DIRBE bar and the Galactic Center - Sun line has been estimated
to be roughly in the range $~\sim 15\gr$ to $~\sim 30\gr$. First
observations of the corresponding angle for the Long bar give an
estimate of $~\sim 40\gr$ \citep{ham00,ben05,lop07,cab08}, while more
recent work favours angles around $25\gr$ to $35\gr$ (Zasowski et al 2011,
private communication). The uncertainties are such that the 
observations could be in agreement with the solution checked here, 
particularly since structures such as spiral arm or ring segments could 
contribute to the observed Long bar signal. In particular, we stress
that in many $N$-body simulations there is often, within the inner
ring a short, leading segment \citep[see
  e.g.][and Fig.~\ref{fig:leading}]{Athanassoula.Misiriotis.02}. 
Furthermore, in many $N$-body
simulations and galaxies 
seen near face-on the boxy/peanut bulge is fatter (i.e. more extended
perpendicular to the bar major axis) than the thin outer part of the
bar. A more accurate answer would  
necessitate a detailed comparison of a large number of simulations to the 
observations. It would be useful to do this work using the Marseille library 
of high-resolution barred galaxy simulations run by one of us (EA).

\subsection{Inter-comparing the various models}
\label{sec:intercompare}

For each model we consider three cases. In the first case, the
non-axisymmetric component describes only the COBE/DIRBE bar or boxy/peanut
bulge located at an angular separation of $20\gr$ with respect to the
Galactic Centre - Sun line. We will refer to this set of models as Case 1. 
From the discussion in the previous section we expect this to have a weaker bar
than it should, since the outer parts, i.e. the contribution of the
Long bar, have been neglected. Nevertheless, we believe it is useful to
consider such cases because a large number of such models have been
applied to our Galaxy for varied purposes. Indeed the five models we
are considering here have been built by other authors for different
purposes.  

In the second case we will include the effect of the Long
bar. Since our approach necessitates the use of an analytic potential,
which has not been calculated so far for objects as complex as the bar
described in Sect.~\ref{sec:1or2bars}, we will model the bar as the
superposition of two bar models, a vertically very thick one which has the
properties of the COBE/DIRBE bar and represents the boxy/peanut bulge,
and a very thin one which represents the Long bar. Trials with
$N$-body simulations show that this is a very reasonable
approximation, no worst than other approximations standardly used in
such modelling. We still need to decide at what angle to place the bar 
major axis with respect to the Galactic Centre - Sun line. Since the 
measurements of the distance of the Long bar are quite accurate, it would be 
tempting to place it at $40\gr$. Simulations, however, show that in many 
cases the end of the bar is not symmetric, but extends considerably further 
towards the leading side, in a form reminiscent of a short arm of ring segment
\citep[e.g.][]{Athanassoula.Misiriotis.02}. This is also seen in a number of
external galaxies. If it is the case for the MW as well, then
observers would be measuring distances both from stars in the thin
outer parts of the bar and from that leading segment and that would make
the bar look as if it were at a somewhat larger angle than 
what it really is. The difference will not be big, maybe 5 or $10\gr$.
For this reason, we have considered many values of the angle between
the bar major axis, but will use for most displays the value of  
$30\gr$, while discussing what the effect of changing this angle is.
Since these two bars are part of the same object, i.e. they rotate together, 
they should have the same pattern speed. To model the Long thin bar, we will 
use the same type of bar as for the thick COBE/DIRBE bar, but with different 
values of the free parameters chosen according to the available observations 
(see Appendix). We will refer to this set of models as Case 2.

As already discussed Sect.~\ref{sec:1or2bars} and also in the previous
paragraph, we believe that the angular separation between the
COBE/DIRBE bar and the Long bar is an artifact due to the
uncertainties of the measurements and the existence of the leading
extension. Nevertheless, in order to follow these observations we will also 
consider the case where the angular separation between bars is of 
$20^{\circ}$ (Case 3). Clearly, since we assume that the 
pattern speeds of the two bars are the same, this situation is dynamically 
unstable due to the forces and torques between the two bars. We nevertheless
discuss it briefly, to follow the observations.

In Table~\ref{tab:forces}, we give a brief summary of the main characteristics 
of each model. For each model (col.~1), we give the type of bar included in 
the potential (col.~2), the Solar radius (col.~3), the value of the pattern 
speed (col.~4), and for each of the three cases, the corotation radius or 
the distance of the equilibrium point $L_1$ to the Galactic Centre, $r_{L_1}$, 
and the three values of bar strength, in sequence $\alpha$, $Q_b$ and 
$Q_{t,L_1}$. The latter are defined as follows:

\begin{itemize}
\item $\alpha$ (cols.~6, 10, and 14) is the ratio of the radial force due 
to the bar's potential to that due to the axisymmetric background, 
evaluated at the solar radius and along the COBE/DIRBE bar major axis. We thus 
have: 

\begin{equation}
\label{eq:qr}
Q_r(r)=\frac{\frac{\partial\Phi_b}{\partial r}\rule[-.25cm]{0cm}{0.5cm}}{\frac{\partial\Phi_0}{\partial r}\rule[0.cm]{0cm}{0.1cm}},
\end{equation}

\noindent and $\alpha=Q_r(R_0)$, where $\Phi_b$ and $\Phi_0$ denote the bar 
and the axisymmetric potential, respectively, and $R_0$ is the Solar
radius. This quantity is analogous to $q_r$ in \citet{ath83}.

\item A similar quantity can be obtained if we use the tangential,
  rather than the radial bar force:

\begin{equation}
Q_t(r)=\frac{\left(\frac{\partial\Phi(r,\theta)}{\partial\theta}\right)_{max}}{r\frac{\partial\Phi_0}{\partial r}},
\end{equation}

\noindent where $\Phi$ is the total potential, 
$\Phi_0$ is its axisymmetric part and the maximum in the numerator is 
calculated over all values of the azimuthal angle $\theta$. The
maximum of $Q_t(r)$ over all radii shorter than the bar extent is
called $Q_b$ (cols.~7, 11, and 15 in Table ~\ref{tab:forces}) and is
often used to measure the bar strength \citep{but03,but04,lau04,but05,dur09,man11}. The radius where this maximum is achieved is defined as $r_{max}$, 
i.e. $Q_b=Q_t(r_{max})$.
 
\item $Q_{t,L_1}$ (cols.~8, 12, and 16) is the value of the tangential force 
at the Lagrange radius or corotation, $Q_{t,L_1}=Q_t(r=r_{L_1})$. This 
indicator was introduced in Paper III and was shown to correlate well with 
morphological features of the galaxy (such as the axial ratio of the rings 
or the pitch angle of spirals).
\end{itemize}

\begin{figure*}
\centering
{\hspace{0.5cm}\Large Case 1} \hspace{5.cm} {\Large Case 2}\hspace{5.cm} {\Large Case 3} \\
\includegraphics[scale=0.23,angle=-90.]{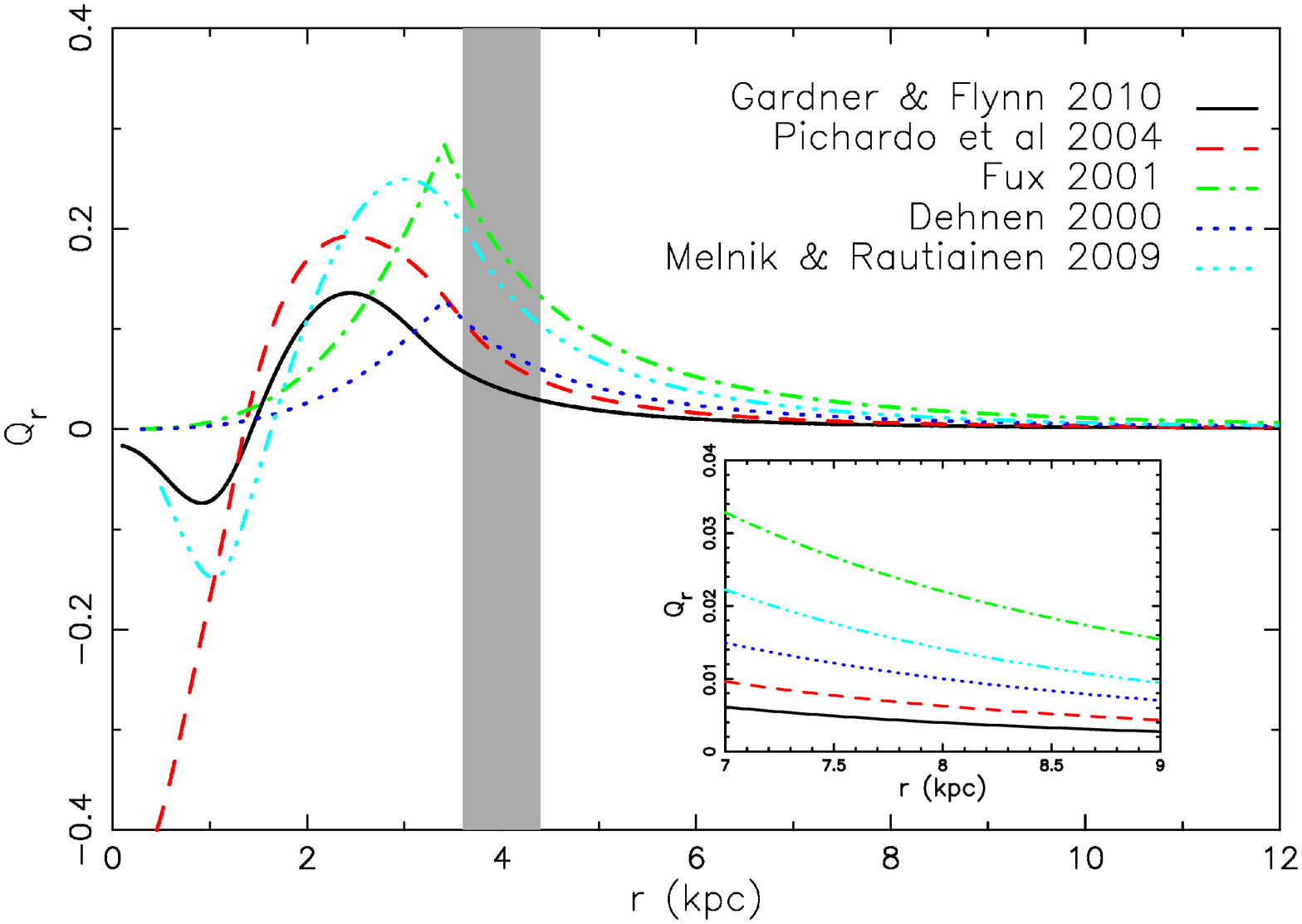}
\hspace{0.15cm}
\includegraphics[scale=0.23,angle=-90.]{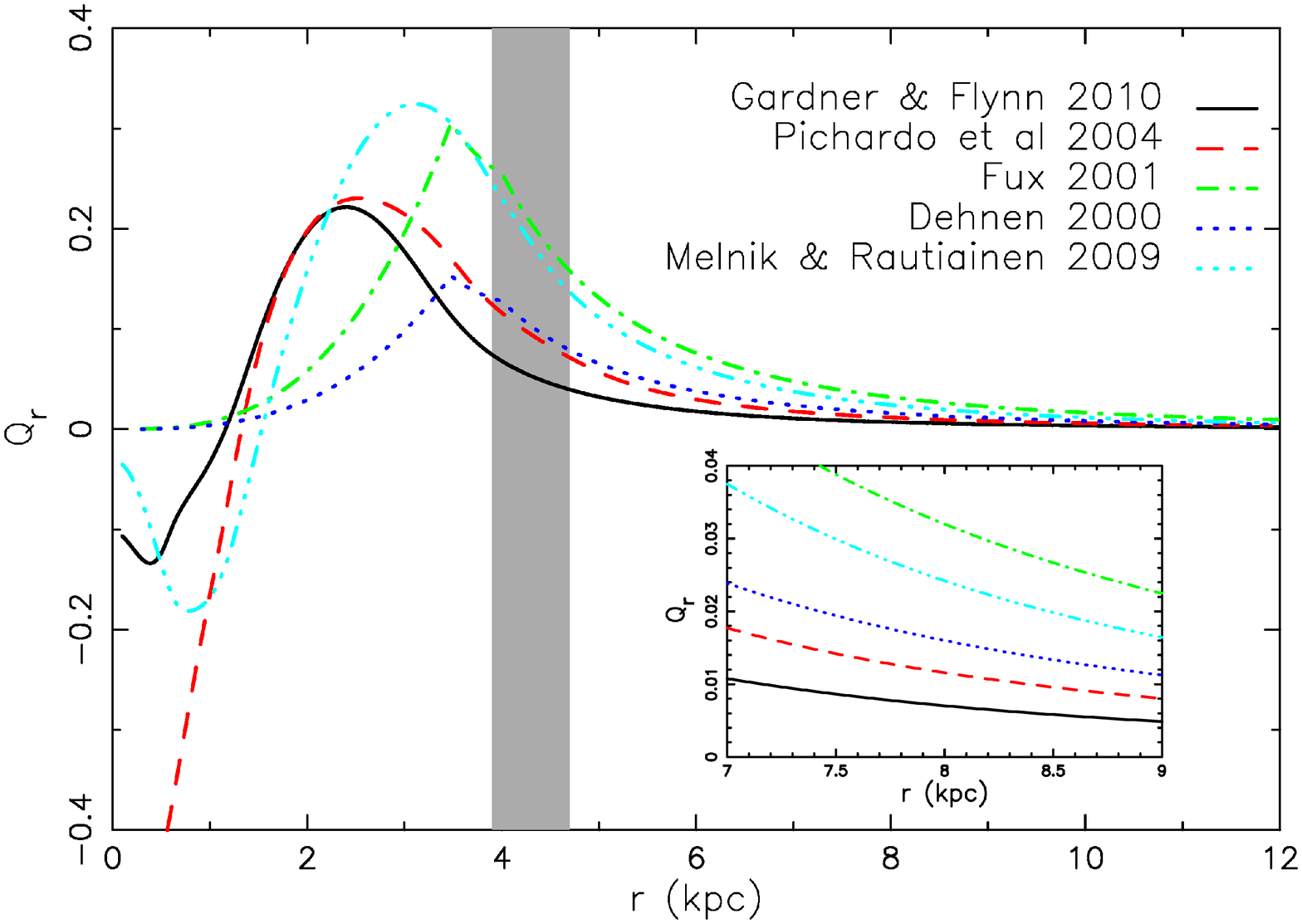}
\hspace{0.15cm}
\includegraphics[scale=0.23,angle=-90.]{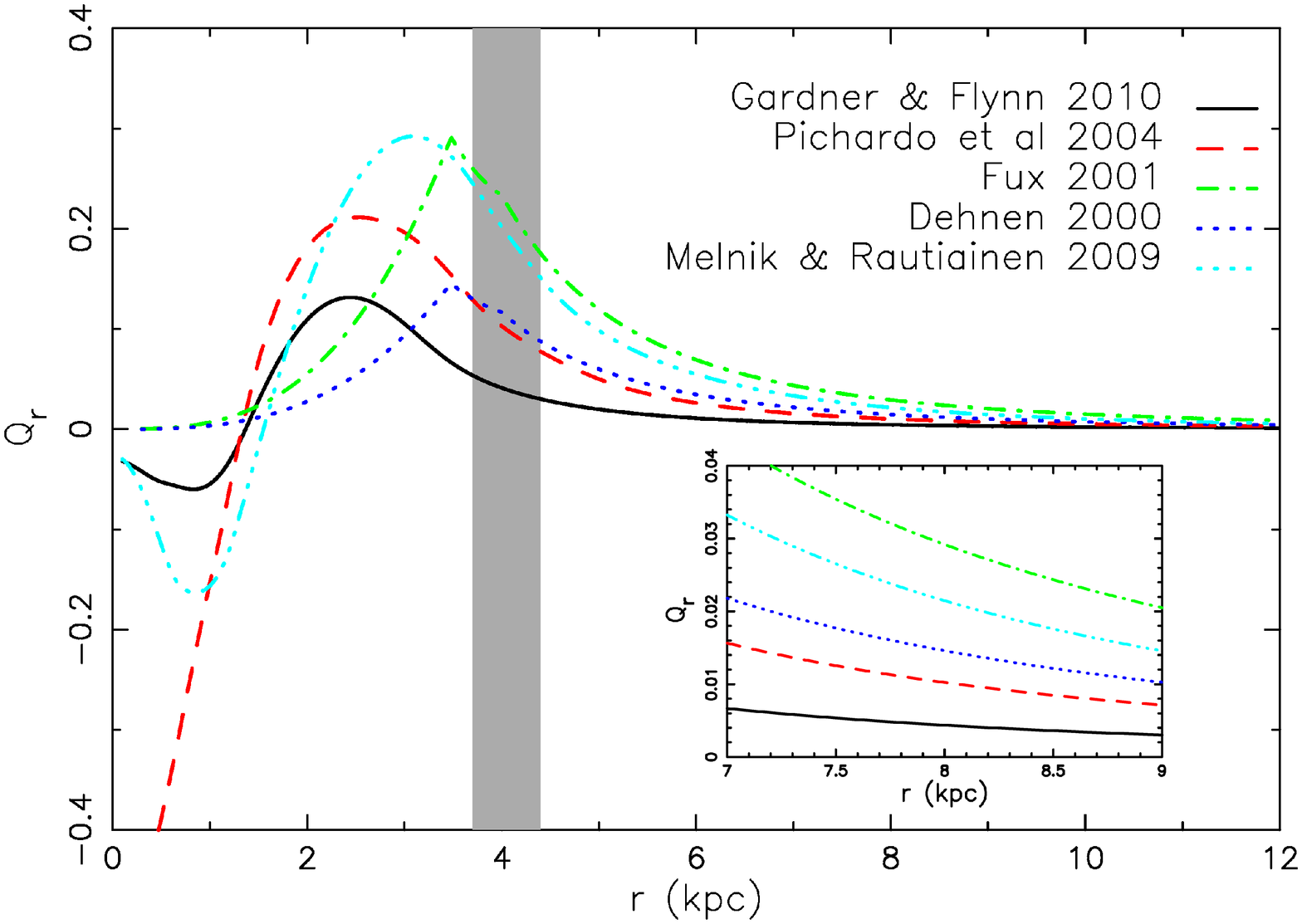}\\
\vspace{0.5cm}
\includegraphics[scale=0.23,angle=-90.]{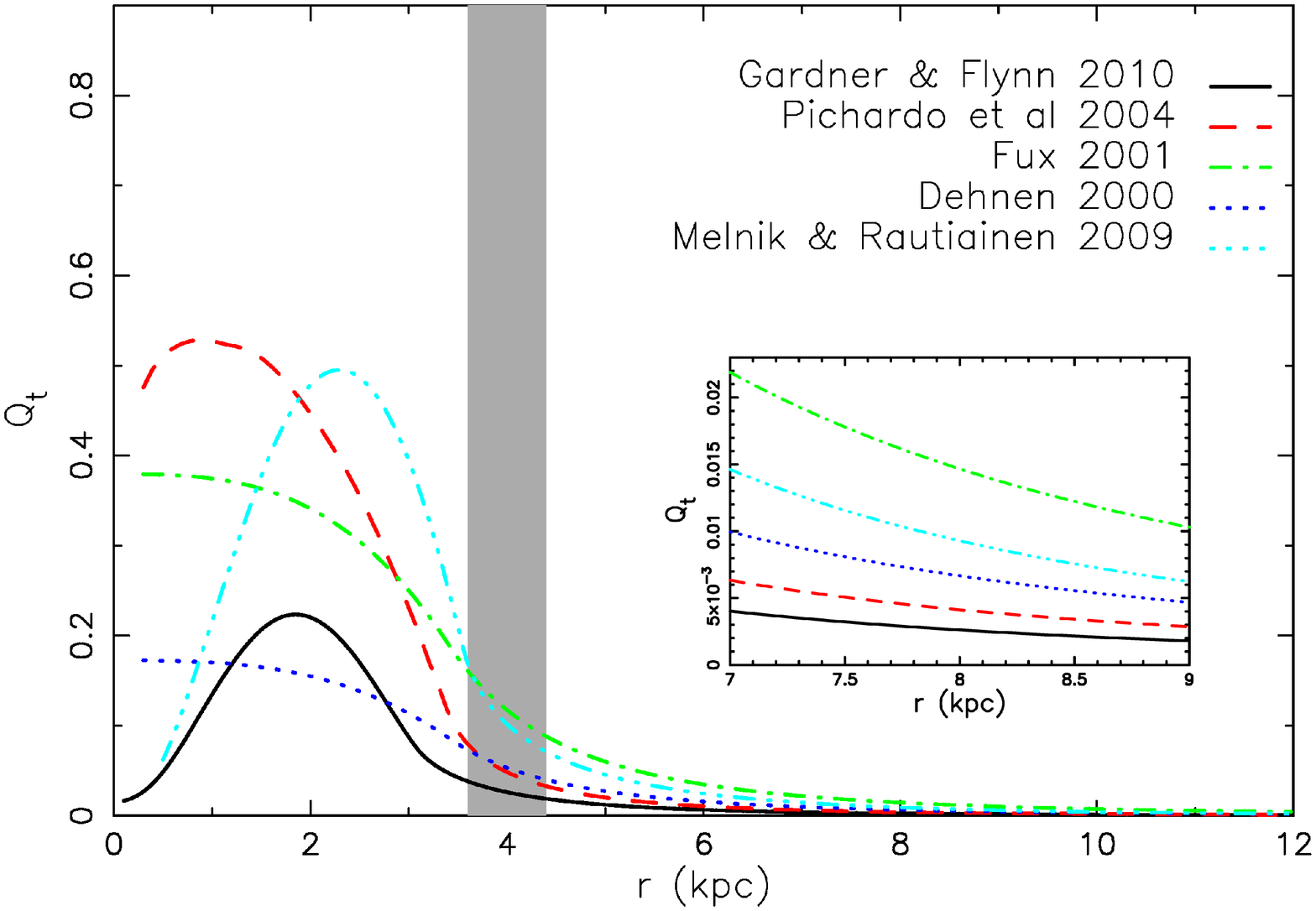}
\hspace{0.15cm}
\includegraphics[scale=0.23,angle=-90.]{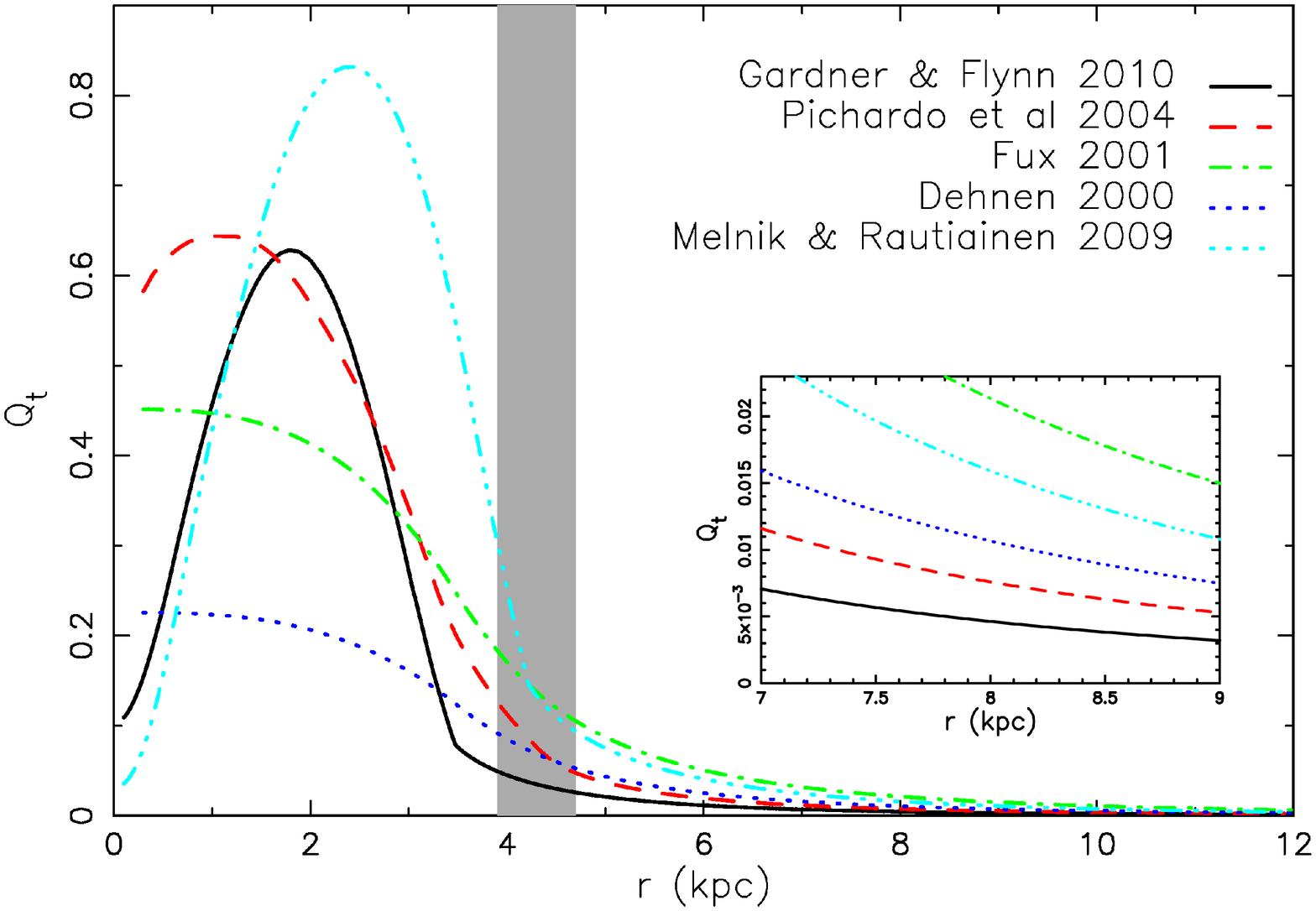}
\hspace{0.15cm}
\includegraphics[scale=0.23,angle=-90.]{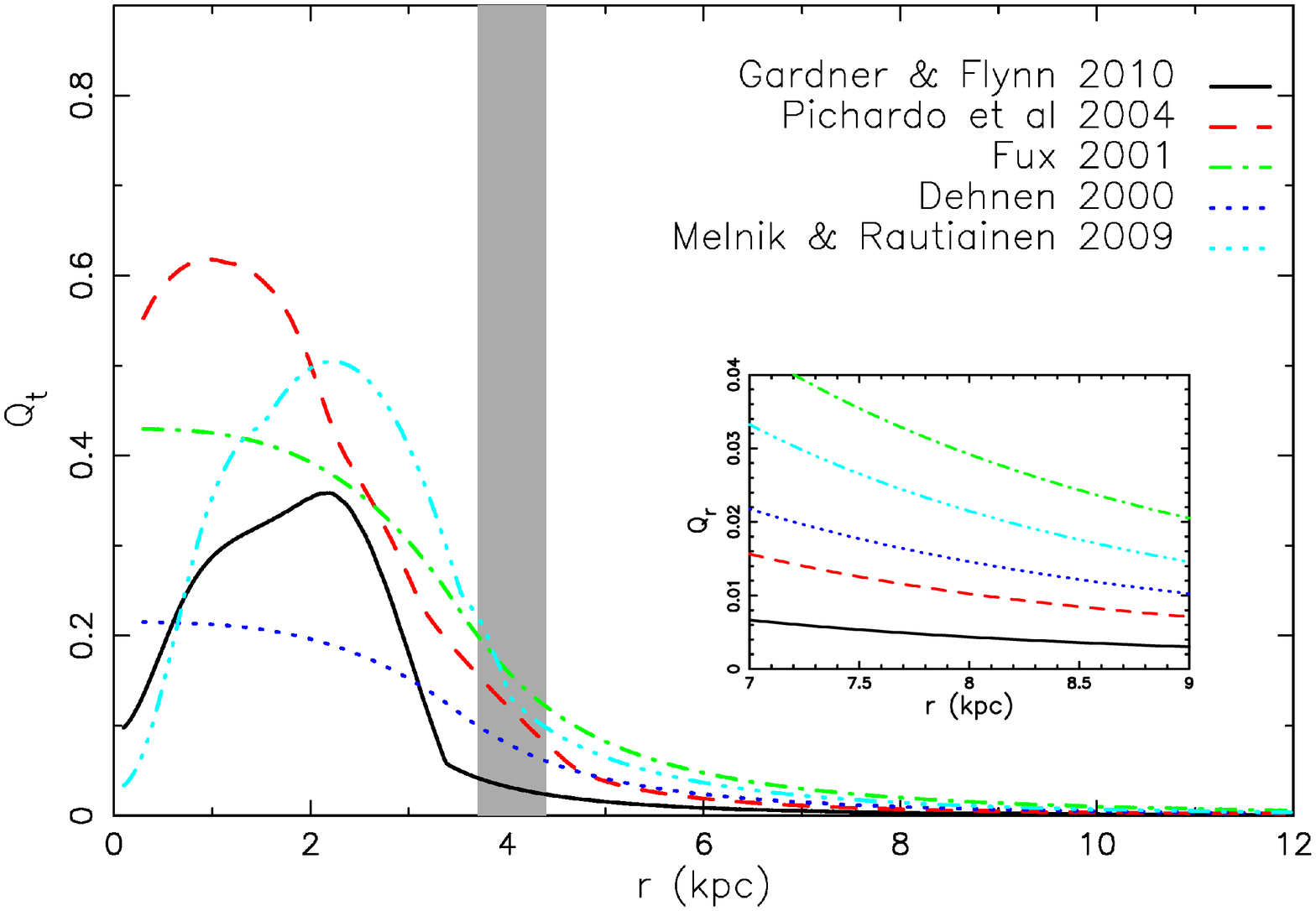}
\caption{Radial profile of $Q_r$ (upper panels) and $Q_t$ (bottom panels) for 
the models used here. Left column: Case 1 with only the COBE/DIRBE bar; middle column: 
Case 2 with the COBE/DIRBE and Long bars aligned;
right column: Case 3 with the COBE/DIRBE and Long bars at an angular separation of $20^{\circ}$. The inlays show 
the Solar neighbourhood region in better resolution. In all panels, the gray 
strip marks the corotation region (see text). }
\label{fig:forces}
\end{figure*}

Note that the values of $\alpha$ and $Q_{t,L_1}$ are evaluated at the outer 
parts of the bar or the disc, whereas $Q_b$ gives the tangential force at a 
radius that lies inside the bar. 
 
In the left panels of Fig.~\ref{fig:forces}, we show $Q_r(r)$ (top panel) 
and $Q_t(r)$ (bottom panel) for the five models with only the 
COBE/DIRBE bar, i.e. Case 1. In the inlet in each panel, we make a zoom of the 
region around the Solar radius. The gray strip marks the range between the 
minimum ($3.6\,kpc$ for PMM04) and the maximum ($4.4\,kpc$ 
for MR09) corotation radius. From the curves in the left panels of
Fig.~\ref{fig:forces}, and  
also from the value of $Q_{t,L_1}$, we note that model Fux01 has the strongest 
bar around corotation, followed, in sequence by MR09, PMM04, Dehnen00, and 
finally GF10. Around the solar radius, both $\alpha$ and $Q_t$
agree in the sequence, which from strongest to weakest is Fux01, MR09, 
Dehnen00, PMM04, GF10. For models MR09, PMM04 and GF10, the region of corotation
is displaced slightly outwards by the introduction of the Long bar,
but it is not affected by any change in the angular separation
between the two bars, as expected. Indeed this displacement is due to the
mass of the Long bar and is independent of its orientation. In
models Fux01 and Dehnen00 the bar is only modelled as a non
axisymmetric forcing, thus introducing no extra mass and no displacement
of corotation outwards. Comparing the curves 
of the middle and right panels of Fig.~\ref{fig:forces}, at the corotation 
and solar regions, we note that the sequence of the models as a function of bar
strength does not change. Note, however, how the introduction of the Long bar 
increases both $Q_r$ and $Q_t$,
in all models and specially in PMM04, GF10 and MR09, which have a bar that comes
from a density distribution. 

Note that the value of $Q_b$ only reflects the behaviour of the tangential 
force in the inner parts of the bar. If we were interested in this region, we 
would use the value of $Q_b$ as the measure of bar strength, and the 
sequence would be as follows: PMM04, followed by MR09, Fux01, GF10, and 
Dehnen00. In Case 3, the introduction of the Long bar does not change the 
sequence, although when the two bars are aligned, Case 2, the
sequence, from the highest to the lowest bar strength, is MR09, GF10, PMM04,
Fux01 and Dehnen00. 

Since we are not interested in the innermost parts of the Galaxy, but
rather in a region around corotation, we will use $Q_{t,L_1}$ as the measure
of bar strength.

\begin{table*}
\begin{center}
\begin{tabular}{|c|c|c|c|c|c|c|c|c|c|c|c|c|c|c|c|}
\hline
\hline
 & & & &\multicolumn{4}{|c|}{Case 1} & \multicolumn{4}{|c|}{Case 2} & \multicolumn{4}{|c|}{Case 3}\\ 
Model & Bar & $R_0$&$\Omega_b$&$r_{L_1}$ & $\alpha$ &$Q_b$ & $Q_{t,L_1}$ & $r_{L_1}$ & $\alpha$ & $Q_{b}$ &  $Q_{t,L_1}$& $r_{L_1}$ &$\alpha$ & $Q_{b}$ &  $Q_{t,L_1}$\\
\hline
\hline
Fux01 & Quadrupole & 8. & 51. & 4.3 & 0.022 & 0.38 & 0.08 & 4.7 & 0.03 & 0.45 & 0.1 & 4.7 & 0.03 & 0.43 & 0.1 \\
\hline
 MR09 & Ferrers & 7.1 & 54. & 4.4 & 0.02  & 0.49 & 0.08 & 4.7 & 0.02 & 0.83 & 0.09 & 4.7 & 0.02 & 0.50 & 0.09 \\ 
\hline
PMM04 & Composite & 8.5 & 60. & 3.6 & 0.005 & 0.53 & 0.08 & 3.9 & 0.01 & 0.61 & 0.12 & 3.9 & 0.008 & 0.54 & 0.13\\
\hline
Dehnen00 & Quadrupole & 8. & 51. & 4.3 & 0.01  & 0.17 & 0.04 & 4.5 & 0.016 & 0.23 & 0.06  & 4.5 & 0.015 & 0.21 & 0.05\\
\hline
 GF10 & Ferrers & 8.5 & 56. & 4.2 & 0.003 & 0.22 & 0.02 & 4.5 & 0.006 & 0.63 & 0.03 & 4.4 & 0.004 & 0.36 &  0.02 \\
\hline
\hline 
\end{tabular}
\caption{Characteristics of the five analytical models. In the first and
second columns, we write the name of the model and the type of bar, 
respectively. In the third column, give the value of the Solar radius and
in the fourth column, we give the value of the pattern speed.
Then for each case, namely Case 1 (potential with only the COBE/DIRBE bar), 
Case 2 (potential with the COBE/DIRBE and Long bar aligned), and Case 3 
(potential with the COBE/DIRBE and Long bar at $20\gr$ of angular separation), 
we give the corotation radius ($r_{L_1}$), the radial force at the Solar 
position ($\alpha$), the maximum of the tangential force ($Q_b$) and the 
tangential force at the corotation radius ($Q_{t,L_1}$).
The units are $kpc$ for distance and $\ksp$ for the pattern speed. }
\label{tab:forces}
\end{center}
\end{table*}

\section{The invariant manifolds}
\label{sec:motion}

The models presented in the previous section are composed of an axisymmetric
component and a non-axisymmetric component, the latter described by either a 
single
bar (Case 1) or two-bars (Cases 2 and 3), rotating clockwise with a constant 
angular velocity, $\Omega_b$. In Cases 2 and 3, both bars are assumed
to have the same pattern speed. We will work in the reference frame where the 
bar is at rest and we use the convention that in this frame, the COBE/DIRBE 
bar is along the $x$-axis. We concentrate on the motion on the $z=0$ plane 
(the Galactic equatorial plane). A full 3D study, albeit for a simple 
logarithmic potential, was treated in \citet{rom09} and revealed that
the motion in the vertical direction can be essentially described by an 
uncoupled harmonic oscillator, whose amplitude is relatively small
and that the 3D structures do not affect the motion in the $z=0$ plane.

\begin{figure}
\centering
\includegraphics[scale=0.4,angle=-90.]{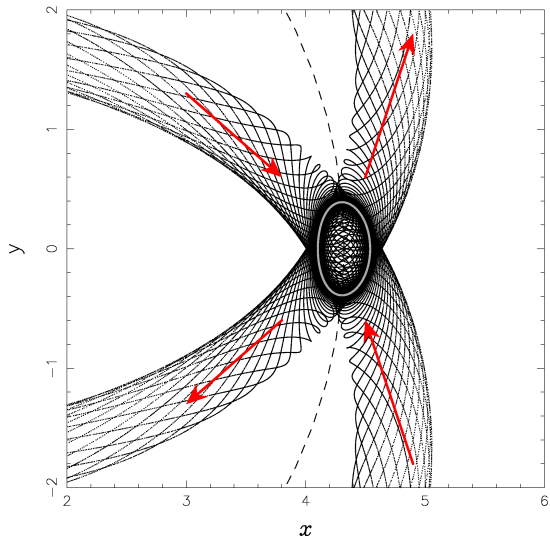}
\caption{Dynamics around $L_1$. The four branches of the invariant manifolds 
(black curves) associated to the Lyapunov orbit (gray curve). The red arrows 
show the sense of circulation, while the black dashed line marks the position of
the corotation radius. }
\label{fig:circulation}
\end{figure}

Our theory is largely based on the dynamics of the Lagrangian points $L_1$ 
and $L_2$ of a two-dimensional galaxy system. These are located where the 
first derivatives of the effective potential vanish, along the bar semi-major
axis, and are unstable saddle points \citep{bin08}.
Each of them is surrounded by a family of periodic orbits, called
Lyapunov orbits \citep{lya49}. Since these orbits are unstable they
cannot trap around them quasi-periodic orbits of the same energy,
so that any orbit in their immediate vicinity (in phase space) will
have to escape the neighbourhood of the corresponding Lagrangian
point. Not all departure directions are, however, possible. The direction
in which the orbit escapes is set by what we call the invariant
manifolds. These can be thought of as tubes that guide the motion
of particles of the same energy as the manifolds \citep{koo00,gom04}.
In Fig.~\ref{fig:circulation}, we show that from each Lyapunov orbit
(light gray thin curve roughly in the middle of the panel), emanate four 
branches: two of them inside corotation (inner branches) and two of them 
outside it (outer branches). Along two of these branches (one inner and one 
outer) the mean motion is towards the region of the Lagrangian point (stable 
manifolds), while along the other two it is away from it (unstable manifolds). 
In this theory the stars travel along the orbits trapped by the manifolds.

We need to stress that the terms ``stable'' and ``unstable'' do not mean that 
the orbits that follow them are stable and unstable, respectively. In fact 
all the orbits that follow the manifold are chaotic, but they are in a loose 
way ``confined'' by the manifolds, so that they stay together, at least for 
a few pattern rotations, in what could be called a bundle. The terms stable 
and unstable refer to the sense of the motion and are related to the saddle
behaviour of the equilibrium points. We propose that these manifolds
and orbits are the building  
blocks of the spirals and rings (Paper II, Paper III). These manifolds do not 
exist for all values of the energy, but only for energies for which the 
corresponding Lyapunov periodic orbit is unstable. This means energies within 
a range starting from the energy of the $L_1$ or $L_2$ ($E_{J,L_1}$) and 
extending over a region whose extent depends on the potential 
\citep{sko02a}.

The morphological and kinematical characteristics of the manifolds in
external barred galaxies are studied in papers III-V. Note that in
these papers, the non-axisymmetric component consists of a single bar. 
The main results relevant to this study can be summarised as follows:

\begin{itemize} 
\item The large-scale structure of the galaxy is related to the
  strength and the pattern speed of the bar  (Paper III). We find that
weak non-axisymmetric perturbations produce manifolds of $rR_1$ ring shape
(an inner ring elongated around the bar and an outer ring whose principal axis 
is perpendicular to that of the bar), while strong non-axisymmetric 
perturbations produce spiral arms or other types of rings. 

\item The range for the shape and size of both the inner and outer ring given
by the models agree with that from observations, where the axial ratio of the 
inner ring spreads uniformly within the range $0.6-0.95$, while the axial 
ratio of the outer ring falls within the range of $0.7-1$ \citep{but86}. 
Furthermore, we find a strong anti-correlation between the axial ratio of the 
rings and the bar strength. 
\item The default number of spiral arms given by the manifolds is two, since 
they are associated with the number of saddle points of the model. The typical 
orientation is trailing, i.e. following the unstable branch of the manifolds, 
and their shape reproduces the characteristic arm winding often observed in 
external galaxies, that is, the spiral arms first unwinds and then
returns to the bar region.  
\item The formation of the manifolds, and therefore of the rings and
spirals, depends on the existence of the saddle Lagrangian points. These
appear when we perturb the axisymmetric potential with a non-axisymmetric
component. We have applied this theory to barred galaxies, but the 
saddle Lagrangian points can be due to other non-axisymmetric perturbations,
such as spiral perturbations.
\item In potentials with a strong m=4 component of the forcing, 
manifolds can also account for four-armed spirals.
\item We also studied the behaviour of collisional manifolds using a
simple model in which the particles within a manifold lose energy
(Papers III and V). We compared collisional and collisionless manifolds
and found that they have very similar shapes, at least for rings and
for the first half turn of spirals. The collisional rings, however are
thinner, i.e. more concentrated, than their collisionless counterparts.
\item Bars are known to evolve secularly, becoming longer and stronger
while slowing down. As a result, the Lagrangian points move outwards
and the $Q_{t,L_1}$ increase. This has important implications for 
manifolds, whose shape and location will change adiabatically with
time. Furthermore it will bring material to the Lagrangian points, 
replenishing the mass reservoirs that fill the manifolds. In the case
of rings, once material is trapped in the manifolds, it can stay there
indefinitely, if the model is stationary, while any increase in the bar
strength can trap more material into the manifolds. 
\end{itemize}

\section{The inner part of the disc}
\label{sec:inner}
Here we study the morphology and kinematics of the models mentioned in 
Sect.~\ref{sec:modelpot}. To compare them morphologically, in 
Sect.~\ref{sec:morphology} we show the invariant manifolds, ``tubes'' that 
guide the motion of the stars, of a given energy level. To study the 
kinematics, in Sect.~\ref{sec:kinematics}, we use the orbits trapped by the 
manifolds, since these are the ones that really trace the kinematics. 
In this case, and in order to make the plots clear and clean, we will 
consider the orbits of minimum energy, since for higher energies, the orbits 
overlap and the area they outline grows thicker (Paper I). Finally, in
Sect.~\ref{sec:MW}, we  
discuss how the manifolds reproduce the observables of the inner part of
the Galaxy, i.e. the region between $2-6\,kpc$, using the up-to-date
MW potentials.

\subsection{The morphology}
\label{sec:morphology}
\begin{figure}
\centering
Composite bar\\
\includegraphics[scale=0.35,angle=-90.]{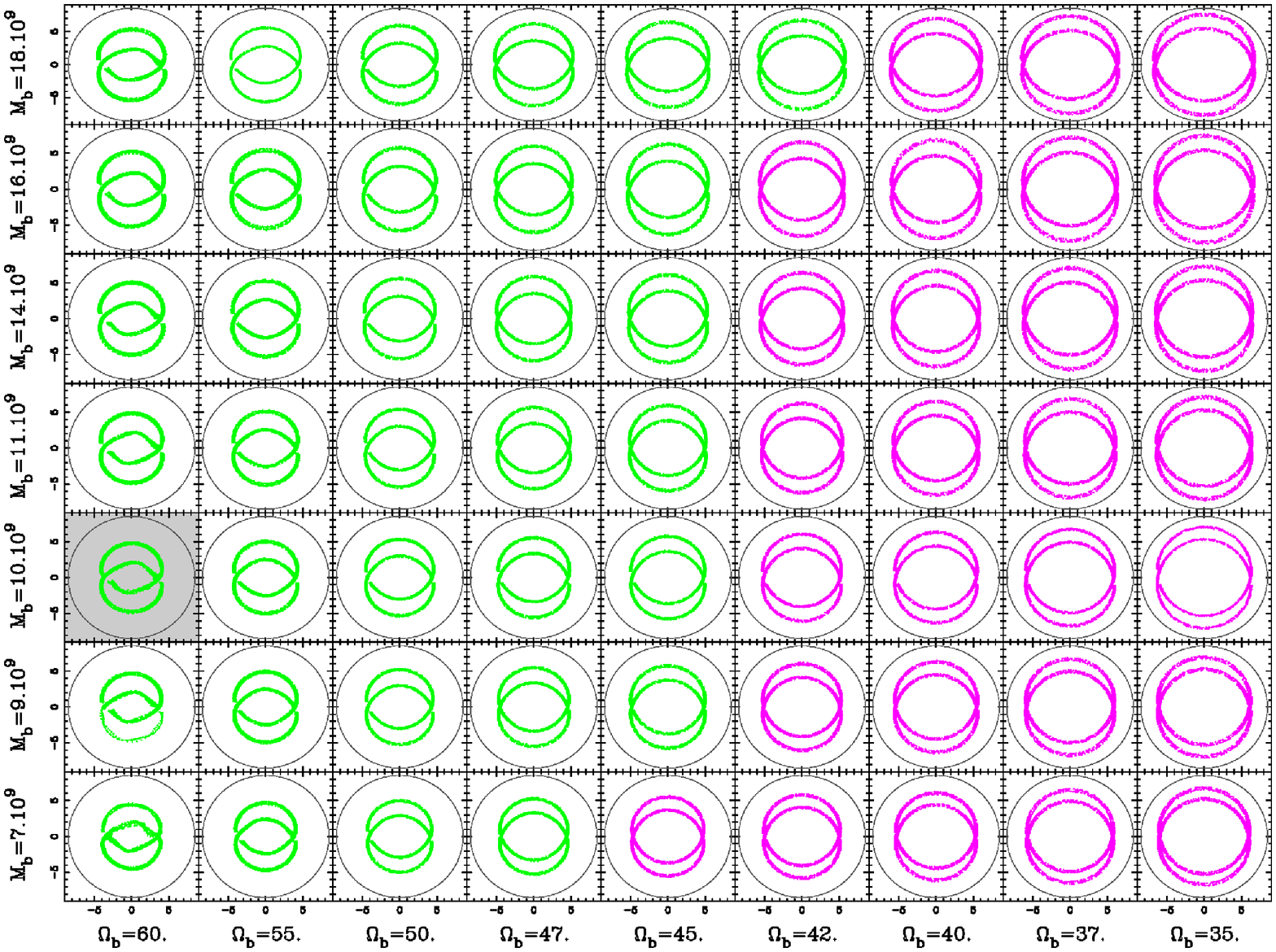}\\
\vspace{0.5cm}
 Quadrupole bar \\
\includegraphics[scale=0.35,angle=-90.]{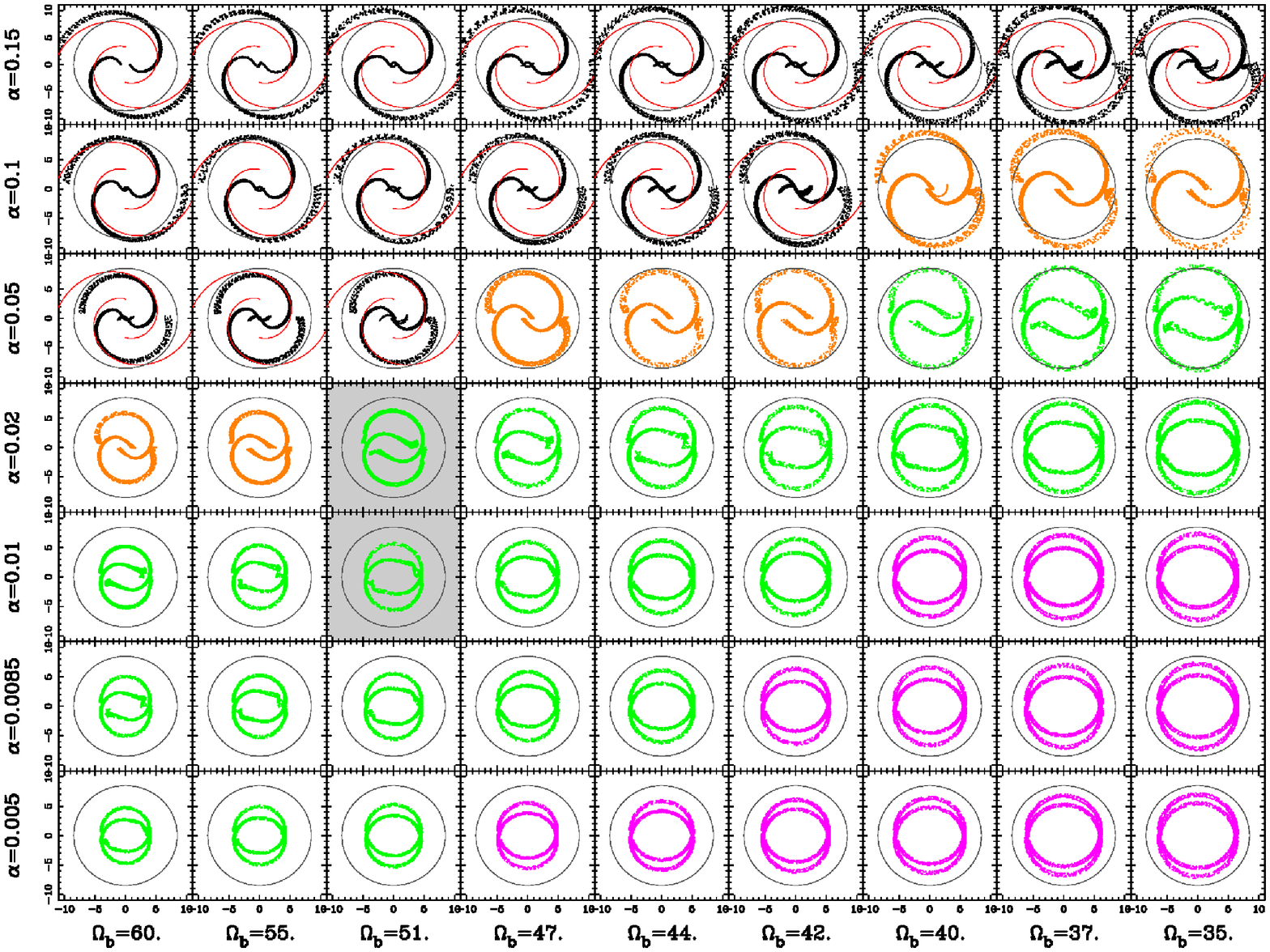}
\caption{2D parameter study of the Composite and Quadrupole bar models for 
Case 1. On the x-axis we decrease the pattern speed from left to right. On 
the y-axis we increase the bar strength from bottom to top, from a bar mass 
$M_b=7\times 10^9 \,M_{\odot}$ to $M_b=18\times 10^9 \,M_{\odot}$ for the 
Composite bar or from $\alpha=0.005$ to $\alpha=0.15$ for the Quadrupole bar. 
The thin black circle in each panel marks the Solar radius.
Top panel: the composite bar model. The panel with gray background corresponds 
to model PMM04. Bottom panel: the quadrupole bar model. The panels with gray 
background correspond to model Fux01 (top: $\alpha=0.02$ and $\Omega_b=51.$) 
and Dehnen00 (bottom: $\alpha=0.01$ and $\Omega_b=51.$). }
\label{fig:2Dstudy}
\end{figure}

\begin{figure*}
{\Large \hspace{1.25cm}Fux01\hspace{2.25cm}MR09\hspace{2.cm}PMM04\hspace{1.75cm}Dehnen00\hspace{2.1cm}GF10}
\centering
\includegraphics[scale=0.9,angle=-90.]{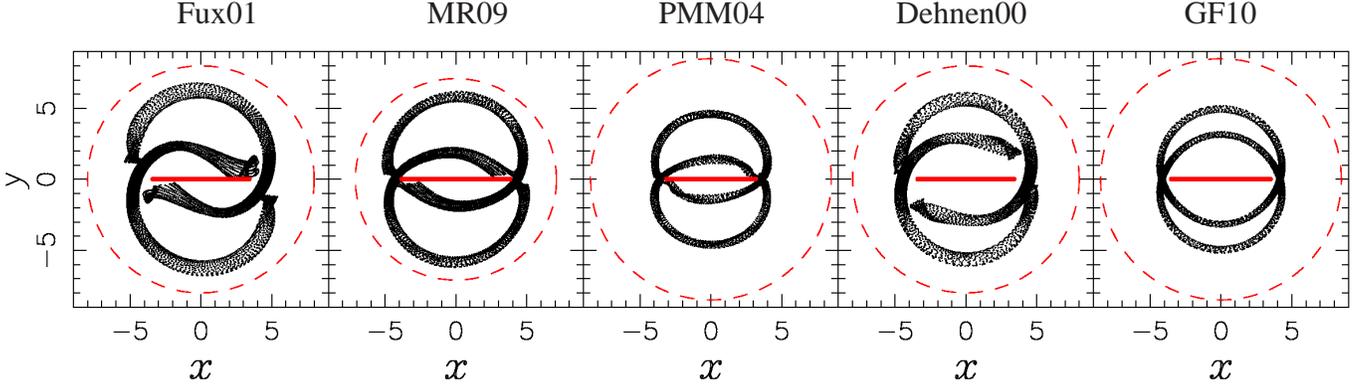}
\caption{Invariant manifolds for the five selected bar potentials and Case 1. 
From left to right: Fux01, MR09, Dehnen00, PMM04, and GF10.
The thick solid red line marks the position and length of the bar. In the case 
of the ad-hoc potentials, it marks the bar scale-length. The thin dashed red 
circle marks the solar radius adopted by the respective authors.}
\label{fig:plot5p}
\end{figure*}

Figure~\ref{fig:2Dstudy} illustrates how the invariant manifolds change as a 
function of the pattern speed and the bar mass / amplitude for two of the 
bar types considered here, namely the composite bar (top panel) and the ad-hoc 
quadrupole bar (bottom panel). The results for the Ferrers bar are similar to 
the composite bar and can be seen by comparing with Fig.~4 of Paper
III. Note that only the COBE/DIRBE bar is modelled here (Case 1) and that as we
increase the pattern speed of a model, the  
Lagrangian radius moves inwards, and the manifolds become attached to the bar. 
To better understand the plots, we give the ratio of the Lagrangian radius over
the bar scale-length. For the composite bar models, the bar scale-length is
fixed to $a=3.13\,kpc$ (see Appendix) and the Lagrangian radius is varied, 
therefore, the ratio $r_L/a$ increases from left to right from $1.47$ to $2$. 
On the other
hand, for the Quadrupole bar models, the ratio $r_L/a$ is fixed to $1.25$, so
the bar length-scale is different in each panel. In the latter case, the ratio 
is inside the range determined by hydrodynamic simulations 
$r_L/a=1.2\pm 0.2$ \citep{ath92}, while in the former, it is outside 
it. We can not, however, extend our search to smaller $r_L/a$ values
because this would entail too big pattern speed values \citep{ger10}. 

In Fig.~\ref{fig:plot5p} we plot the invariant manifolds for all
Case 1 models that we consider here, that is, up-to-date MW models
with only the COBE/DIRBE bar and we use  
the set of parameters that the authors consider to describe best the MW disc 
potential. In each panel we plot the invariant manifolds  
of a given energy. The morphology does not depend on the energy value chosen, 
provided this is near the Lagrangian point energy (Paper I)
\footnote{The energy chosen for each model is: $|(E_{J,Fux01}-E_{J,L_1})/E_{J,L_1}|=6.6\times 10^{-4}$, $|(E_{J,MR09}-E_{J,L_1})/E_{J,L_1}|=2.5\times 10^{-3}$, $|(E_{J,PMM04}-E_{J,L_1})/E_{J,L_1}|=6.5\times 10^{-4}$, $|(E_{J,Dehnen00}-E_{J,L_1})/E_{J,L_1}|=7.2\times 10^{-3}$,  and $|(E_{J,GF10}-E_{J,L_1})/E_{J,L_1}|=7.2\times 10^{-3}$.}.  
The panels are ordered from left to right, from the strongest bar 
model to the weakest bar model, namely Fux01, MR09, PMM04, Dehnen00 and GF10, 
according to the $Q_{t,L_1}$ measure of the force (see Table~\ref{tab:forces}).  
All cases present both an inner and an outer ring and can be classified as 
$rR_1$ ring morphologies. The axial ratio of the rings, however, are clearly 
different from one model to another. In columns 2-5 of  
Table~\ref{tab:rings} we give the axial ratio and the major diameter 
of the inner and outer rings. Dehnen00 and GF10 have an inner ring 
axial ratio that falls well within the observational range ($0.6-0.95$) for 
the SB galaxies \citep{but86}. In contrast, the inner rings in Fux01, MR09 
and PMM04 are too elongated with axial ratios ranging from $d_i/D_i=0.43-0.5$, 
respectively. In the particular cases of MR09 and PMM04, whose bar potential 
is built from a density distribution, the locus of the inner manifolds falls 
well inside the bar ellipsoid, so that, when self-gravity would be considered, 
the two could merge, as discussed in Paper IV. Therefore, we cannot consider 
the inner rings of MR09 and PMM04 as proper inner rings surrounding the bar and 
this justifies their low axial ratio. The bar in Fux01 is ad-hoc and we cannot 
confirm this fact. For Dehnen00 and GF10, the axial ratio of the inner ring 
increases with decreasing bar strength, from being oval, $d_i/D_i=0.66$ for 
Dehnen00, to more circular, $d_i/D_i=0.77$ for GF10, in good agreement with 
what was found in Paper IV. There are also differences in the size. The inner 
ring in Fux01, MR09 and PMM04 is very elongated, as mentioned above, not 
reaching $3\,kpc$ along the minor axis. On the other hand, Dehnen00 and GF10 
have a major diameter of $D_i=4.5\,kpc$ and $D_i=4.2\,kpc$, respectively.   

The axial ratio of the outer ring in the five models falls well within the 
observational range ($0.7-1$) for SB galaxies \citep{but86} and correlates 
with the bar strength as measured by $Q_{t,L_1}$, as found for other models in 
Paper IV. Fux01, with the 
strongest bar, has the more eccentric outer ring, $d_o/D_o=0.74$, while GF10, 
with the weakest bar, has a more circular ring with $d_o/D_o=0.86$.
The major diameter is around $D_o=6\,kpc$. In Fux01 the outer ring reaches 
$6.9\,kpc$ along the major diameter, and in MR09, and Dehnen00 it is around 
$6\,kpc$, while GF10 and PMM04 have the smallest outer rings with a major
diameter of $5.3\,kpc$ and $5.0\,kpc$, respectively. 

\begin{table*}
\begin{center}
\begin{tabular}{|c|c|c|c|c|c|c|c|c|c|c|c|c|}
\hline
\hline
 & \multicolumn{4}{|c|}{Case 1} & \multicolumn{4}{|c|}{Case 2}& \multicolumn{4}{|c|}{Case 3}   \\
Model & $d_i/D_i$ & $D_i$ & $d_o/D_o$ & $D_o$ & $d_i/D_i$ & $D_i$ & $d_o/D_o$ & $D_o$ & $d_i/D_i$ & $D_i$ & $d_o/D_o$ & $D_o$ \\
\hline
\hline
 Fux01   & 0.43 & 4.6 & 0.74 & 6.9 & 0.31 & 4.7 & 0.68 & 7.2 & 0.31 & 4.6 & 0.74 & 7.1 \\
\hline 
 MR09    & 0.46 & 4.3 & 0.75 & 6.2 & 0.40 & 4.7 & 0.71 & 6.8 & 0.32 & 4.6 & 0.78 & 6.9 \\
\hline
PMM04    & 0.50 & 3.6 & 0.78 & 5.0 & 0.32 & 3.9 & 0.76 & 6.6 & 0.41 & 3.7 & 0.78 & 5.5 \\
\hline
Dehnen00 & 0.66 & 4.5 & 0.80 & 6.1 & 0.54 & 4.5 & 0.76 & 6.3 & 0.50 & 4.5 & 0.82 & 6.3\\
\hline
GF10     & 0.77 & 4.2 & 0.86 & 5.3 & 0.72 & 4.4 & 0.81 & 5.7 & 0.78 & 4.1 & 0.82 & 5.1\\
\hline
\hline
\end{tabular}
\caption{Sizes of the inner and outer rings in the three cases considered, from 
left to right, only the COBE/DIRBE bar (Case 1), COBE/DIRBE and Long bar aligned 
(Case 2) and COBE/DIRBE and Long bar at $\phi=20\gr$ (Case 3). For each of 
the cases and for each of the models, we give, in the first and second columns, 
the axial ratio and the major diameter of the inner ring, $d_i/D_i$ and $D_i$, 
respectively, (in $kpc$), while in the third and fourth columns, we give the 
same values for the outer ring, $d_o/D_o$ and $D_o$, respectively.}
\label{tab:rings}
\end{center}
\end{table*}

In Case 2, where the Long bar is aligned to the COBE/DIRBE bar, the global
morphology is still that of an $rR_1$ ring, but there is an effect on the
shape and kinematics of the rings (see Sect.~\ref{sec:kinematics}). The axial
ratio of the outer ring decreases approximately by $5\%$ in the mean,
compared to that of Case 1, while the axial ratio of the inner ring
decreases on average by $20\%$ (see columns 6-9 of
Table~\ref{tab:rings}). Changing the angle between the bar major axis and the
direction of the Galactic Centre - Sun line makes no difference to
this result. In Case 3, the axial ratio of the outer ring, $d_o/D_o$,
essentially does not change compared to that of Case 2. The inner
ring, on the contrary, gets more elongated with a mean
decrease of the axial ratio, $d_i/D_i$, of about $30\%$ compared to the values
given by Case 1, in agreement with the elongated 3-kpc arm obtained by
\citet{hab06} from the distribution of maser stars in the inner MW.
As discussed in Paper IV this can be due to the fact
that we include orbits that form the outer parts of the bar. The size
of the global structure is similar to that of the
single bar model (see columns 10-13 of Table~\ref{tab:rings}). This
means that there is in general an agreement with what is observed in
external galaxies.

\subsection{The kinematics}
\label{sec:kinematics}
\begin{figure}
\includegraphics[scale=0.45]{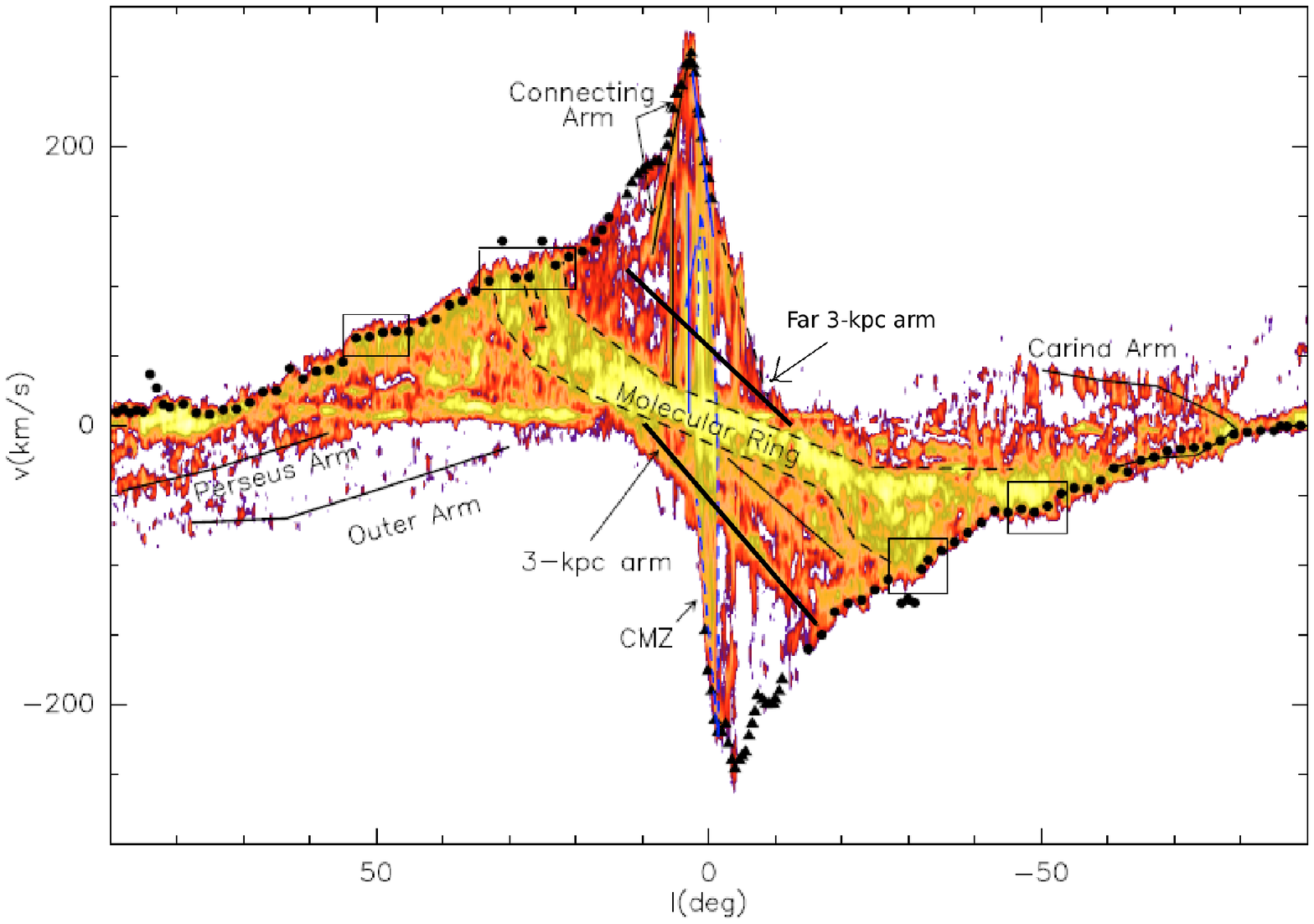}
\caption{The ($l$,$v$) diagram obtained from the observations of
  \citep{dam01} and \citep{dam08}, as plotted by \citep{rod08}.
The line-of-sight velocity is given with respect to the Local Standard of Rest.
The solid lines trace the position of some remarkable features such as the
locus of the spiral arms, the 3-kpc arm and the Connecting Arm. The black
dashed lines indicate the contour of the GMR. The solid circles are the
terminal velocities measurements of \citet{fic89} using CO, while the
triangles are the terminal velocities determined from the HI data of 
\citet{bur93}. The boxes mark the position of the Sagittarius, Scutum, Norma
and Centaurus tangent points, located, respectively, at $l\sim 50^{\circ}$,
$l\sim 30^{\circ}$, $l\sim -30^{\circ}$, and $l\sim -50^{\circ}$.}
\label{fig:COobs}
\end{figure}

This section is devoted to the analysis of the kinematics provided by the 
manifolds in the selected models. Fig.~\ref{fig:COobs} shows the 
line-of-sight heliocentric velocity given at each galactic longitude, 
hereafter ($l$,$v$) diagram, obtained from CO observations \citep{dam01}. 
The main features found in the inner $2-6\,kpc$ of the Galaxy are marked with 
solid and dashed lines, namely the near and far 3-kpc arms and the over-density 
crossing the ($l$,$v$) at $l = 0\gr$ and $v = 0\,km\,s^{-1}$ (GMR or 
Galactic Molecular Ring). The rest of the features are related to the central 
part of the Galaxy or to the outer arms, so we will not consider them in
our comparisons. Note also that these features were observed in the
CO, while we compare them to collisionless manifolds. This, nevertheless,
is possible since in Papers III and V we showed that the shape of collisional
and collisionless manifolds does not differ much, at least for
rings and for the first half turn of spirals (except for the width,
which is smaller in the collisional cases). Furthermore, in external
galaxies where it is easy to check, we note that the shapes of the
gaseous and stellar rings are very similar, with the notable
difference that the rings in the young stars are thinner than the ones
in the old stars, in good agreement with our calculations. Finally,
in external galaxies there are few, if any, purely gaseous rings with
no stellar component. This would anyway be contrived, since it
would involve a strong concentration of gas which would not form
stars. We can thus proceed with our comparisons.

\begin{figure*}
\begin{center}
\includegraphics[scale=0.7,angle=-90.]{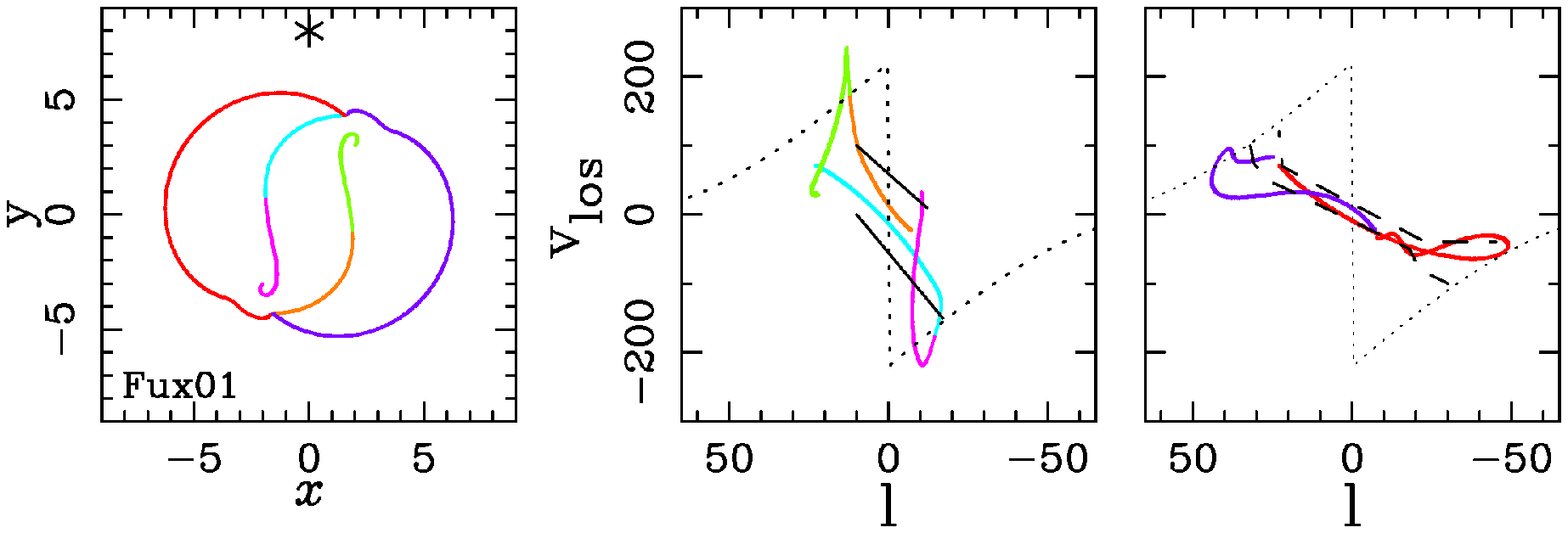}\\ \vspace{0.2cm}
\includegraphics[scale=0.7,angle=-90.]{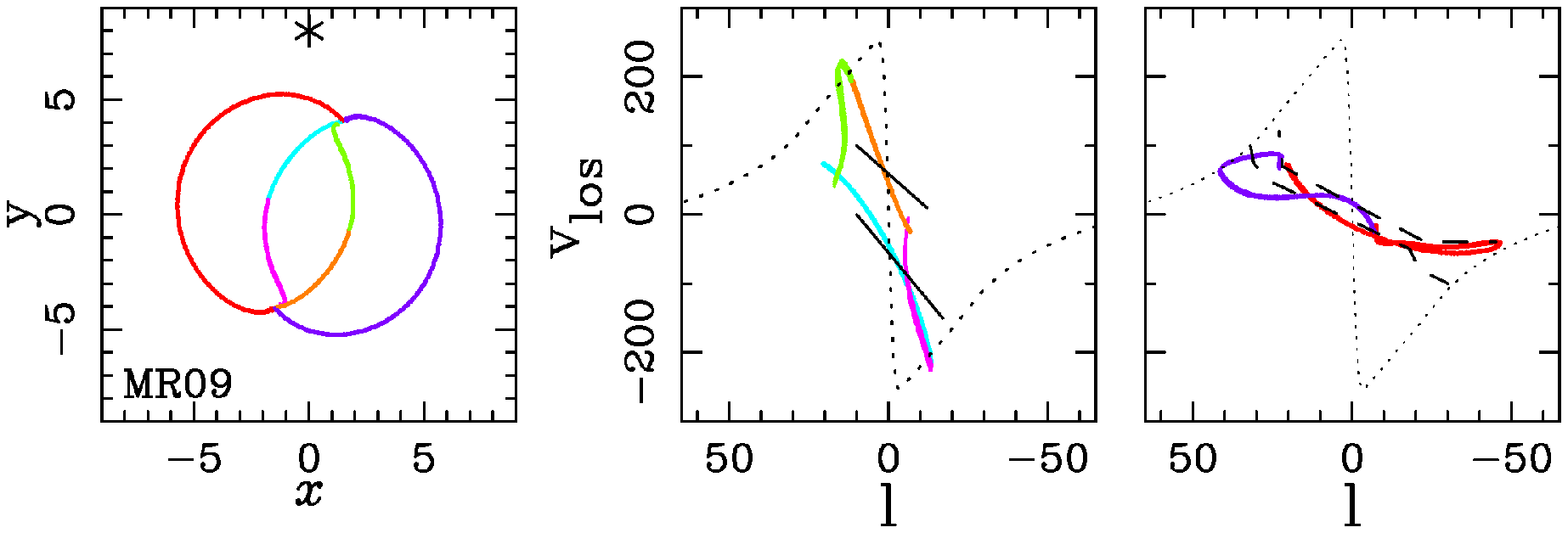}\\ \vspace{0.2cm}
\includegraphics[scale=0.7,angle=-90.]{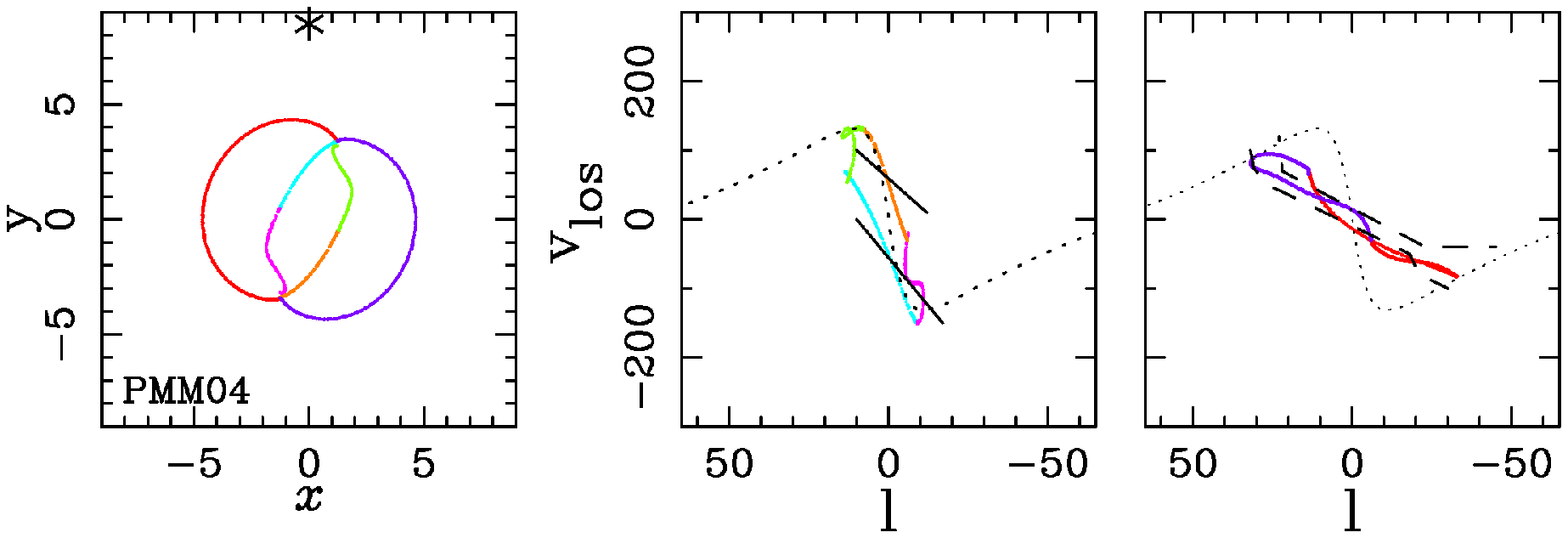}\\ \vspace{0.2cm}\includegraphics[scale=0.7,angle=-90.]{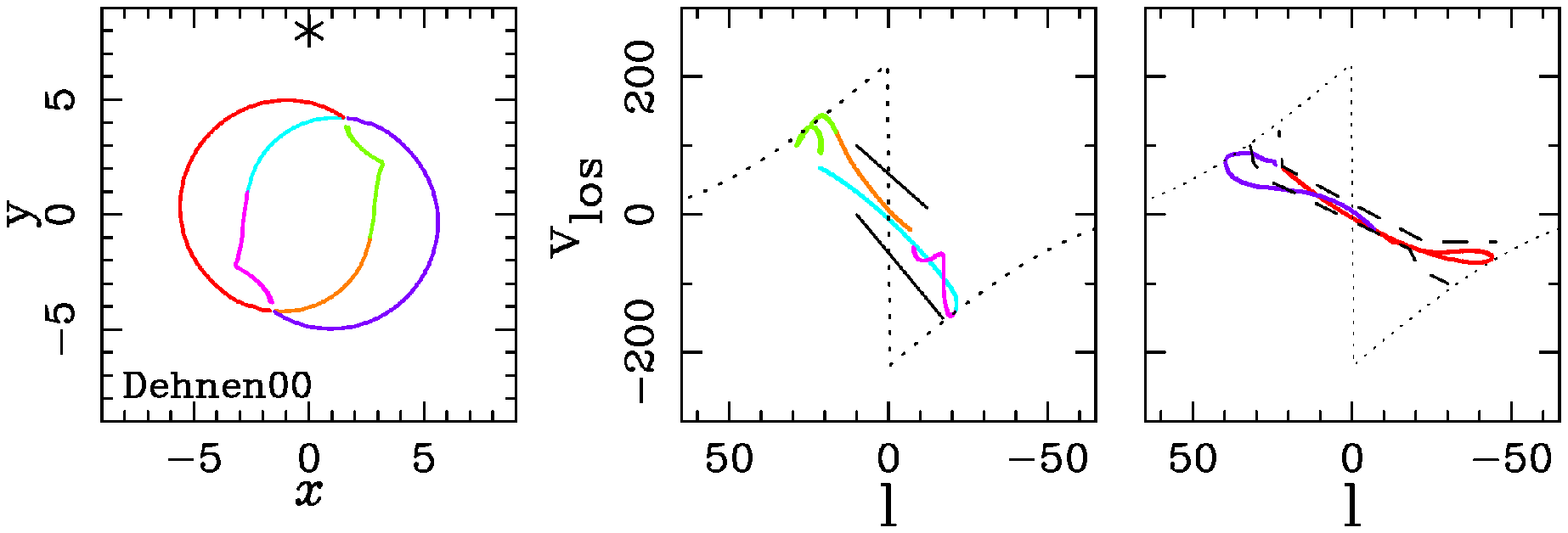}\\ \vspace{0.2cm}
\includegraphics[scale=0.7,angle=-90.]{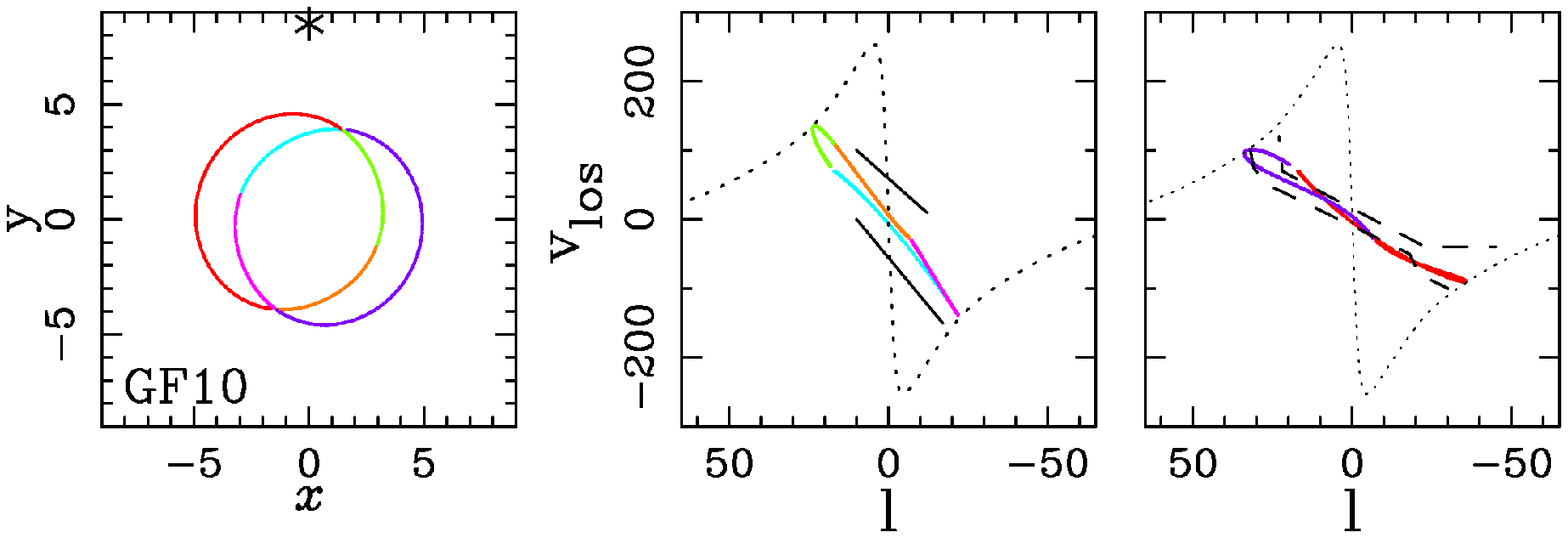}

\end{center}
\caption{($l$,$v$)-diagrams of the five selected models with only the 
COBE/DIRBE bar, i.e. Case 1: from top to bottom, Fux01, MR09, PMM04, Dehnen00 
and GF10 model. Left column: orbits in the (x,y)-plane, the colours showing 
different parts of the rings; Middle and right column: (l,v) diagram
of the inner and outer manifolds, respectively. The circular 
terminal velocity is given by the black dotted line.  }
\label{fig:COinvman}
\end{figure*}

\begin{figure*}
\centering
\includegraphics[scale=0.7,angle=-90.]{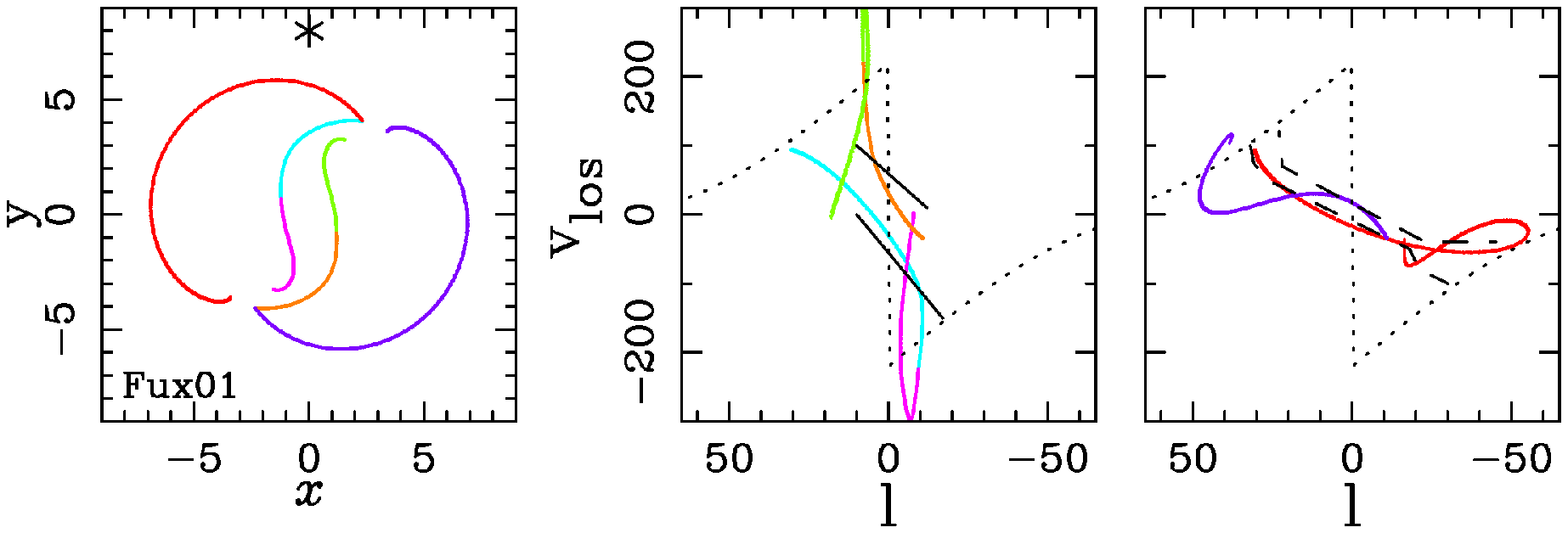}\\ \vspace{0.25cm}
\includegraphics[scale=0.7,angle=-90.]{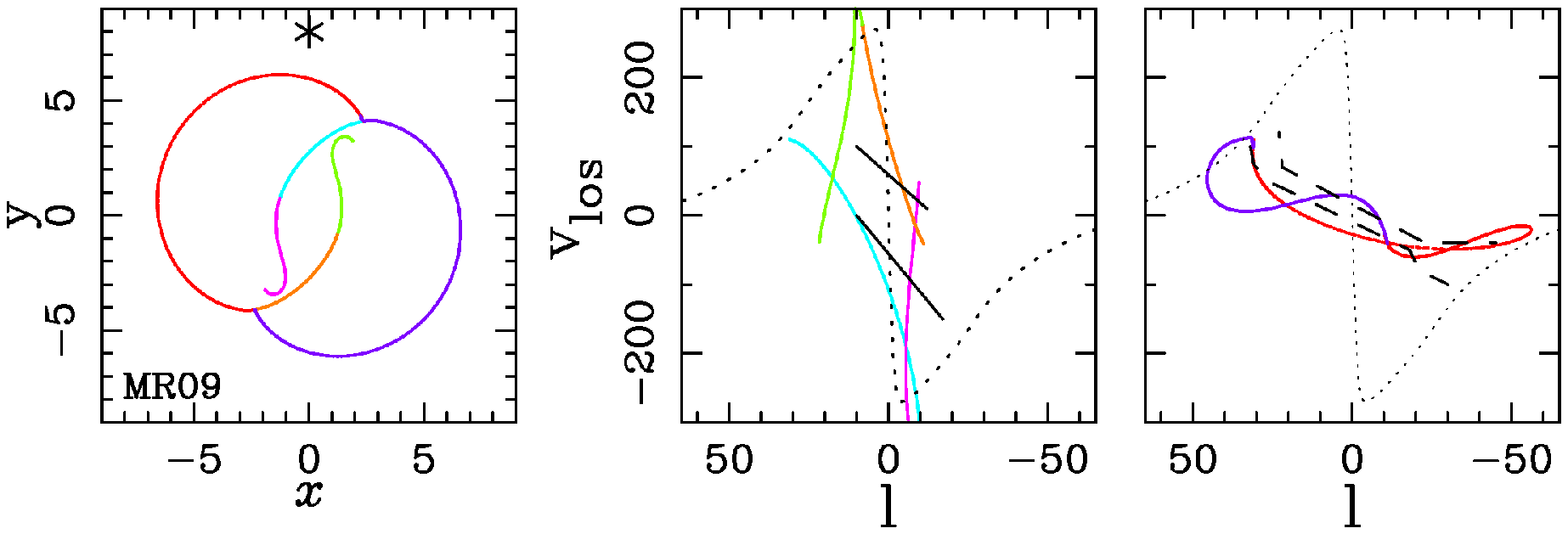}\\ \vspace{0.25cm}
\includegraphics[scale=0.7,angle=-90.]{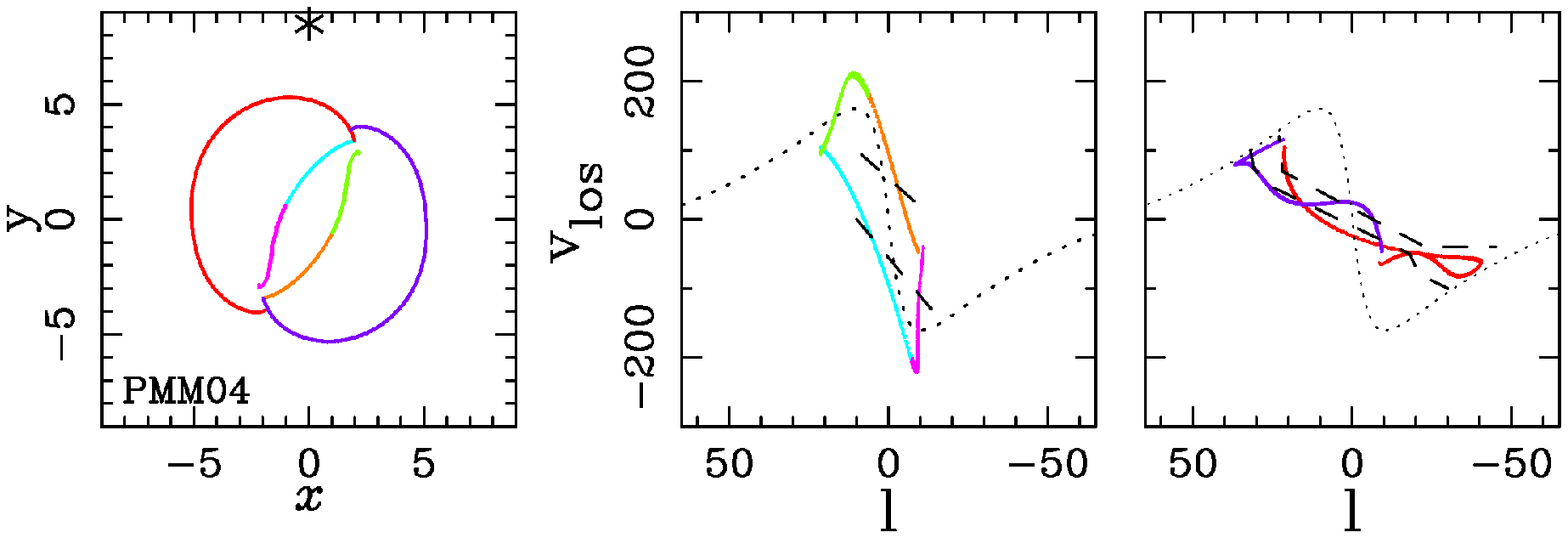}\\ \vspace{0.25cm}
\includegraphics[scale=0.7,angle=-90.]{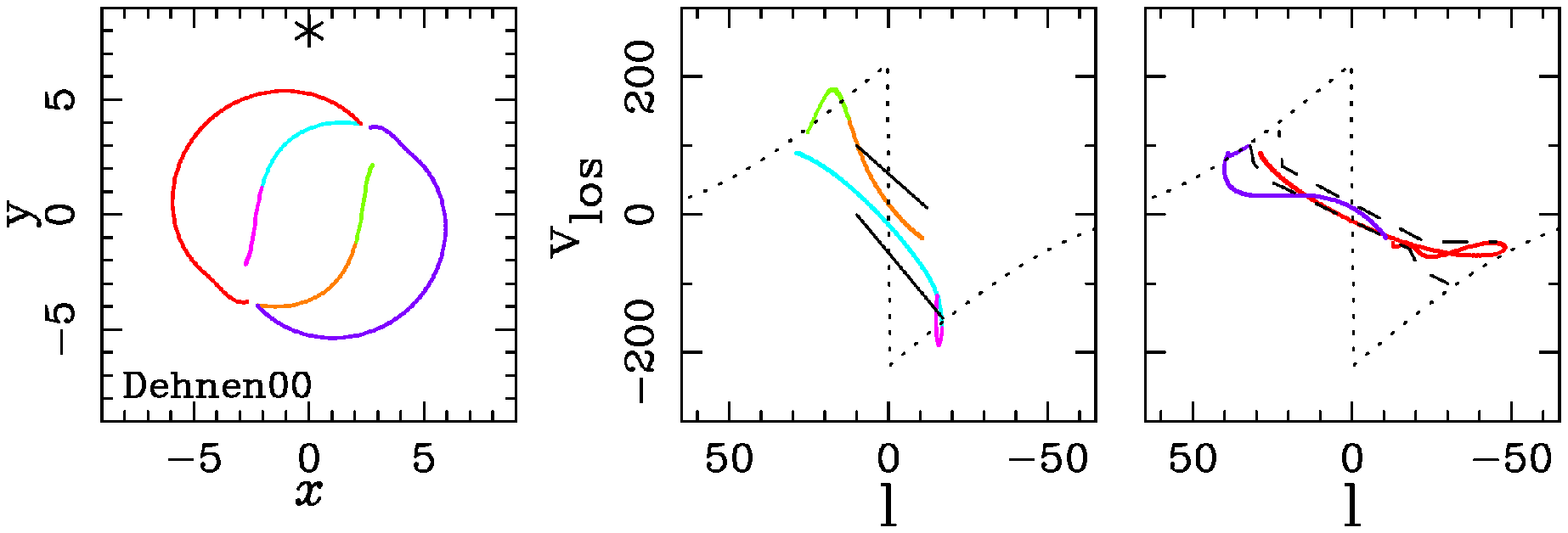}\\ \vspace{0.25cm}
\includegraphics[scale=0.7,angle=-90.]{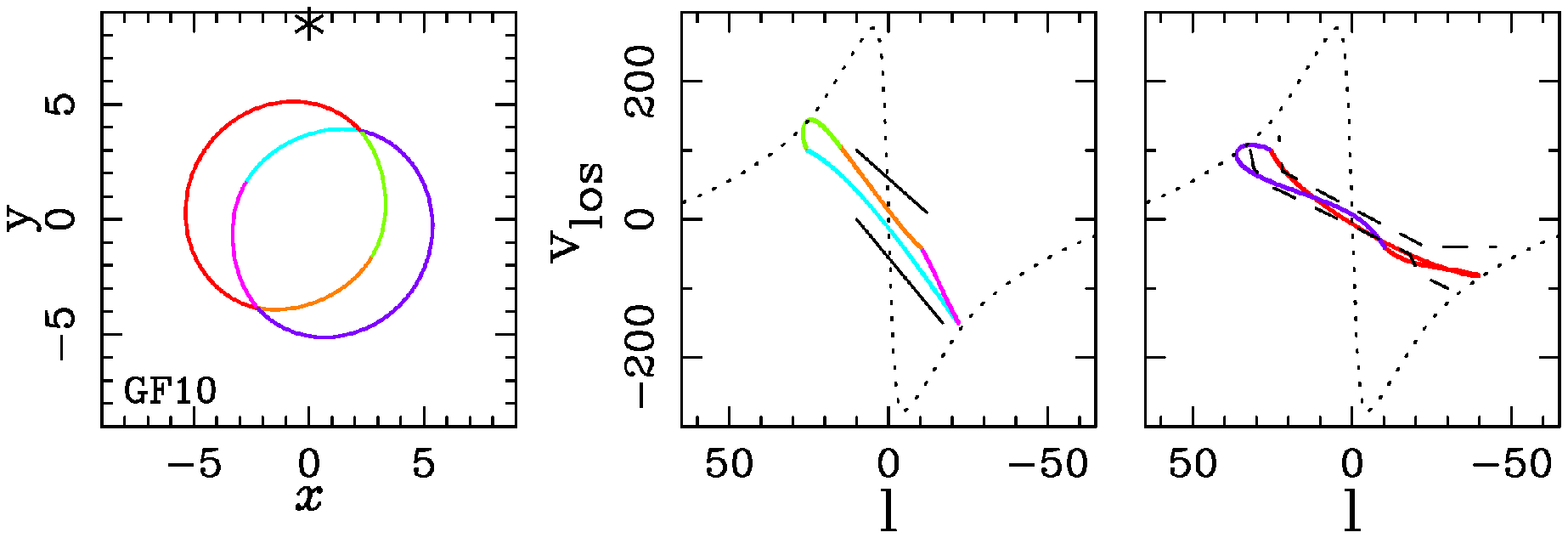}
\caption{As in fig.~\ref{fig:COinvman}, but now for Case 2, where the two bars 
are aligned at $30\gr$ from the Galactic Centre - Sun line.
}
\label{fig:COinvman_2bars_phi0}
\end{figure*}

\begin{figure*}
\centering
\includegraphics[scale=0.7,angle=-90.]{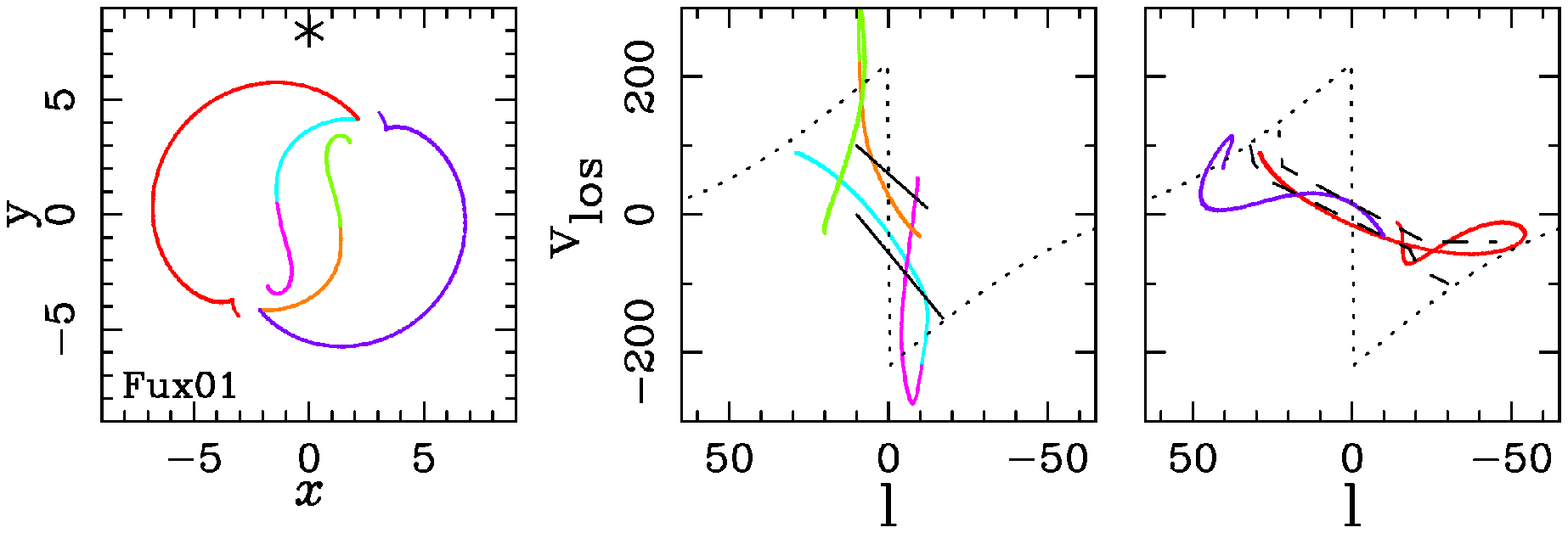}\\ \vspace{0.25cm}
\includegraphics[scale=0.7,angle=-90.]{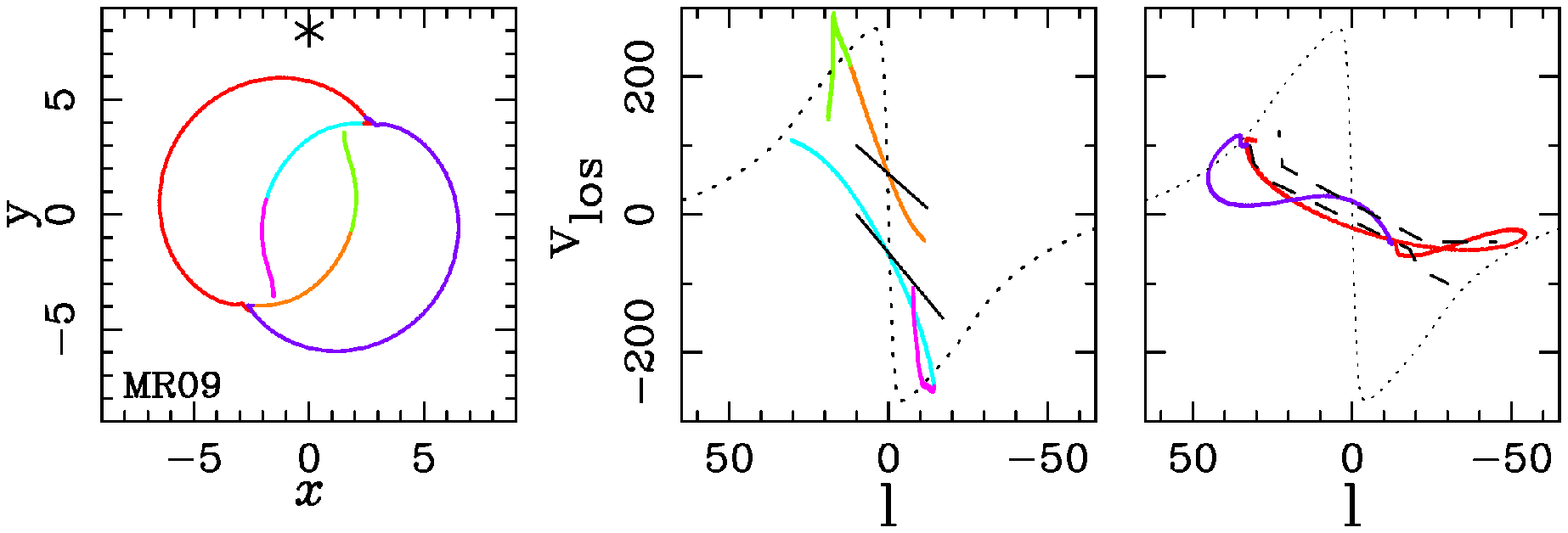}\\ \vspace{0.25cm}
\includegraphics[scale=0.7,angle=-90.]{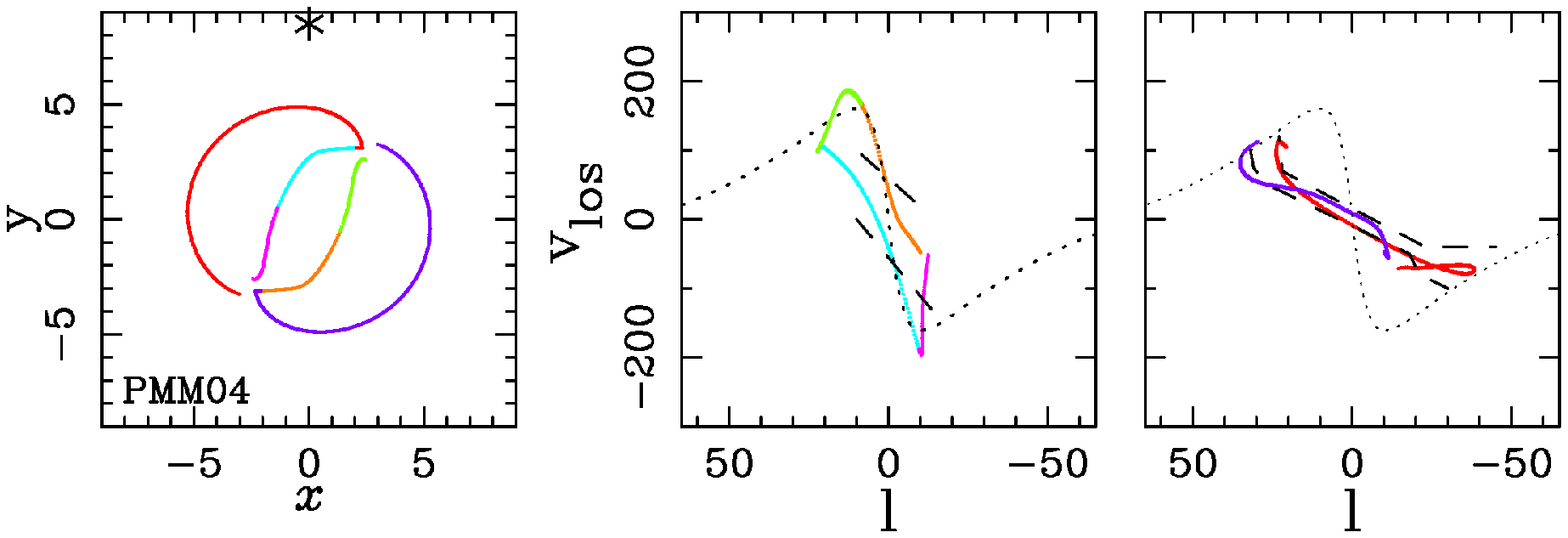}\\ \vspace{0.25cm}
\includegraphics[scale=0.7,angle=-90.]{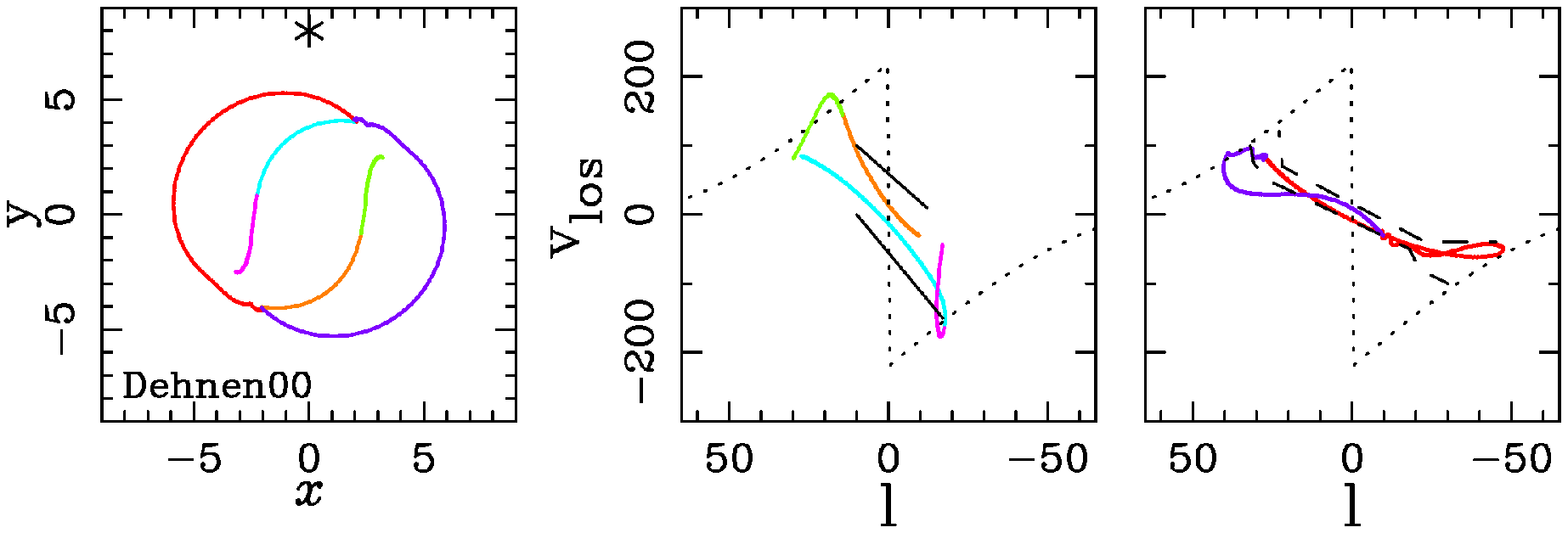}\\ \vspace{0.25cm}
\includegraphics[scale=0.7,angle=-90.]{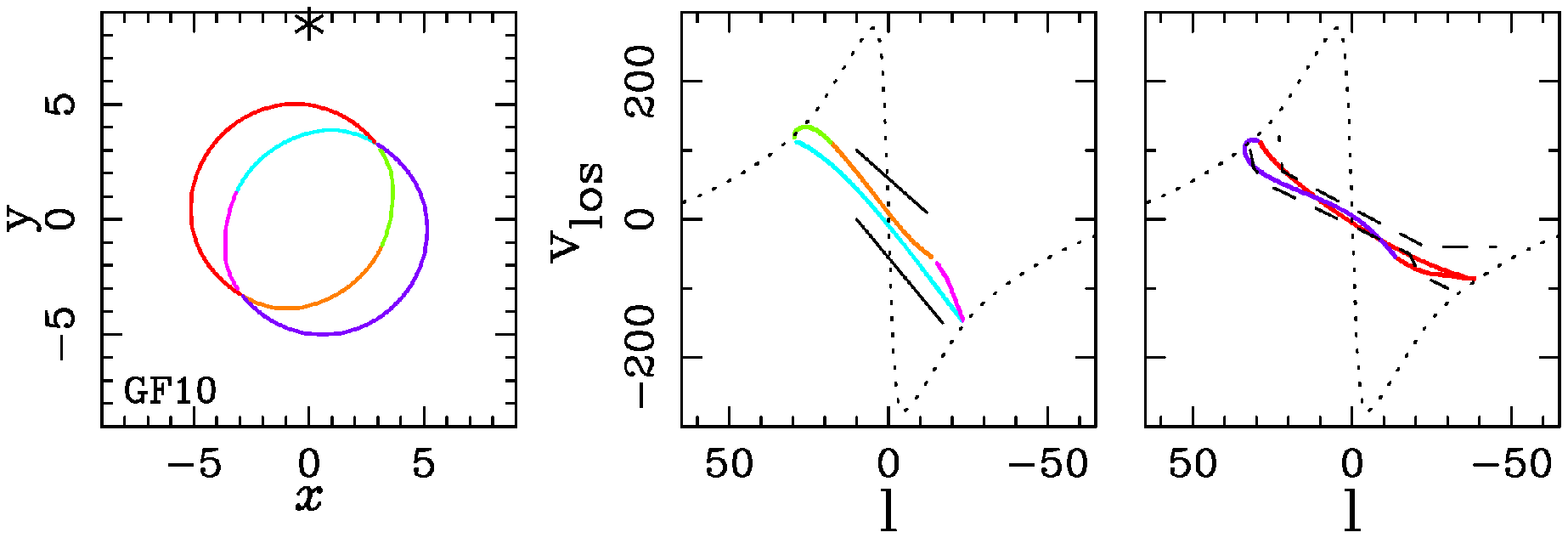}
\caption{As in fig.~\ref{fig:COinvman}, but now for Case 3, i.e. for 
models with two bars with an angular separation of $20^{\circ}$. }
\label{fig:COinvman_2bars}
\end{figure*}

For each model, we compute the orbits trapped by the manifolds with the 
energy closest to the energy of $L_1$ (minimum energy for the manifolds to 
exist), and we show them in the first column of Figures~\ref{fig:COinvman}, 
\ref{fig:COinvman_2bars_phi0} and \ref{fig:COinvman_2bars}, for the COBE/DIRBE 
bar only (Case 1), the two-bar models aligned (Case 2), and the two-bar models 
with and angular separation of $20\gr$ (Case 3), respectively. In the middle 
and right panels of the same Figures, we plot the ($l$,$v$) diagram separately 
for the inner and outer branches of the rings, respectively. The colours mark 
the different parts of the ring. We also show in a thin dotted black line the 
axisymmetric component of the terminal velocity curve, this is, the maximum 
line-of-sight velocity at a given longitude. The black solid and dashed lines 
are extracted from Fig.~\ref{fig:COobs} and trace the position of the near 
and far 3-kpc arm and of the GMR, respectively. 

The kinematic features in the ($l$,$v$)-diagram that should be characterised 
and taken into account for the subsequent comparison with observations
(Sect.~\ref{sec:MW}) are: first, where and by how much the
non-axisymmetric component exceeds the circular terminal velocities; second, the
shape of the lines traced on the   
diagram compared to the over-densities on the observed one; third, the non-zero 
value of the line-of-sight velocity at $l=0\gr$ for these over-densities 
(expansion or contraction in the radial Sun - Centre direction). 

In several models, we observe that the velocities of the orbits in the
manifolds exceed the circular terminal velocities at low
longitudes. For example, 
the inner rings of Fux01 and MR09 in all three cases (one or two bars)
exceed the terminal velocities at low longitudes (see the second column of 
Figs.~\ref{fig:COinvman}, \ref{fig:COinvman_2bars_phi0} and 
\ref{fig:COinvman_2bars}). This is due to the non-axisymmetric
motions, but can in some cases be excessive (see discussion in
Sect.~\ref{sec:MW}). It is also worth mentioning that the introduction of
the Long bar changes the ($l$,$v$) diagram in almost all cases. In Case 2, 
where the Long bar is aligned with the COBE/DIRBE bar, in all models except 
for GF10, the velocities of the inner ring exceed the circular terminal 
velocity curve, due to the excess of mass introduced. In Fux01 and MR09, however, the maximum velocity is well 
above the terminal 
curve. In Case 3, it increases the velocities in models PMM04, Dehnen00 and
GF10 to values slightly exceeding the terminal velocities at certain longitudes.

Table~\ref{tab:vlosl0} summarises the line-of-sight velocities along the 
Galactic centre - Sun line, i.e. $l=0\gr$. We see that the range of variation
for this velocity is $6-50\,km\,s^{-1}$ in absolute value, for the models in
Case 1. The introduction of the Long bar in Case 2 increases the line-of-sight
velocity at $l=0\gr$ in all models and, in Case 3, in all models but PMM04, 
where it slightly decreases. 

\begin{table}
\begin{center}
\begin{tabular}{|c|c|c|c|c|c|c|}
\hline
\hline
 & \multicolumn{2}{|c|}{Case 1} & \multicolumn{2}{|c|}{Case 2}&\multicolumn{2}{|c|}{Case 3}\\
Model & $v_{los}$ N & $v_{los}$ F  & $v_{los}$ N & $v_{los}$ F & $v_{los}$ N & $v_{los}$ F \\
\hline
\hline
 Fux01   & -15. & 13. & -30. & 30. & -27. & 28. \\
\hline 
 MR09    & -46. & 44. & -108. & 109. & -59. & 58.\\ 
\hline
PMM04    & -51. & 50. & -94. & 95. & -42. & 42.\\
\hline
Dehnen00 & -7. & 6. & -15. & 15. & -15. & 13.\\
\hline
GF10     & -6. & 6. & -13. & 12.  & -9. & 9.\\
\hline
\hline
\end{tabular}
\caption{Line-of-sight velocities with respect to the LSR at $l=0\gr$. N stands for the Near 3-kpc arm,
while F stands for the Far 3-kpc arm. Units are in $km\,s^{-1}$.}
\label{tab:vlosl0}
\end{center}
\end{table}

\begin{figure}
\centering
\includegraphics[scale=0.8,angle=-90.]{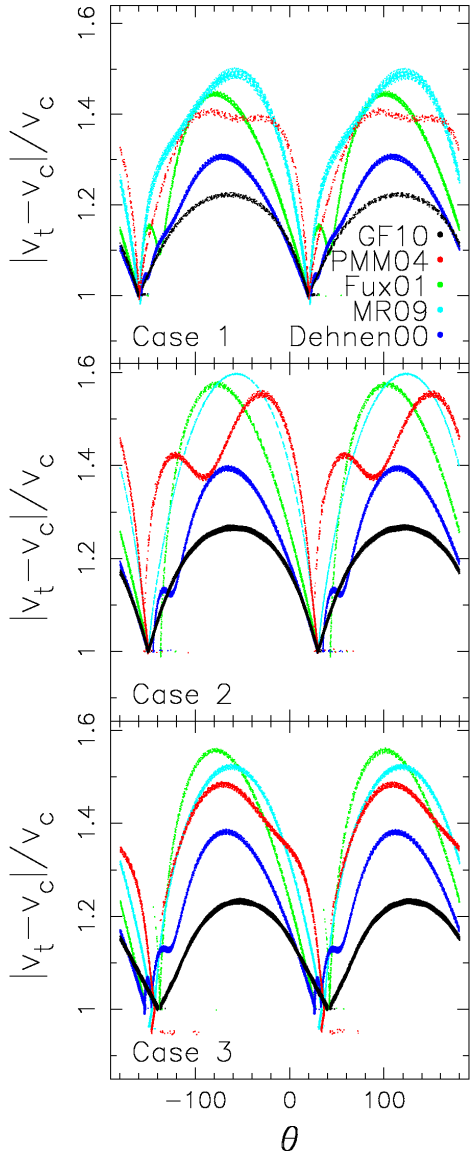}\\
\caption{Relative deviation of the tangential velocities in the inertial 
frame with respect to the circular velocity of the outer ring as a function 
of the azimuthal galactocentric angle $\theta$. Top panel: Case 1,  
models with only the COBE/DIRBE bar. Middle panel: Case 2, two-bar models 
aligned. Bottom panel: Case 3, two-bar models at an angular separation of
$20\gr$. The colours represent the same type of models in all panels.}
\label{fig:vels_MR}
\end{figure}

The kinematics along the outer ring also provide information about the
deviation from the circular velocity curve. Fig.~\ref{fig:vels_MR} shows the 
relative deviation of the tangential velocity in the inertial frame from the 
circular velocity for the orbits in all models with the COBE/DIRBE bar. The 
angle $\theta$ is defined as the azimuthal galactocentric angle with origin 
on the Galactic Centre - Sun line and measured clockwise. The circular 
velocity at a given radius is given by the axisymmetric component in the case 
of Fux01 and Dehnen00 models and by the axisymmetric component plus the 
$m=0$ component of the bar in MR09, GF10 and PMM04 models. Note that in each 
case, the deviation from a flat rotation curve is less than $20\,km\,s^{-1}$, 
in agreement with Fig. 6 of \citet{kra01}.

\subsection{Towards a manifold model of the inner MW}
\label{sec:MW}

In the previous sections we analysed in an exhaustive way the results given by 
the manifolds for a wide set of models including the possibility of modelling
only the COBE/DIRBE bar or both the COBE/DIRBE bar and the Long bar. We claim that
in general the manifolds can interpret the observables both morphologically
and kinematically. We now compare the results with the MW observations
and analyse in detail what manifold models are more likely to 
reproduce the 3-kpc arm and the GMR.

The observed characteristics of the 3-kpc arm are two-fold: it is elongated 
along the bar and it has non-zero velocities at $l=0\gr$  
($-53\,km\,s^{-1}$ for the near arm and $56\,km\,s^{-1}$ for the far arm 
\citep{dam08}, although other studies based on ammonia and water in 
absorption give a value of $-43\,km\,s^{-1}$ \citep{wir10}. This sets
a tentative estimate of the error bar of about $10\,km\,s^{-1}$. Among the 
set of models considered here, we can say that when only the COBE/DIRBE bar
is considered (Case 1) MR09 and PMM04 reproduce this feature 
(see Table~\ref{tab:vlosl0}, whereas Fux01, Dehnen00 and GF10 have absolute
values below $15\,km\,s^{-1}$. 

The introduction of the Long bar, either aligned or misaligned, in 
the models makes the inner ring more elongated and increases the 
velocities along $l=0\gr$, suggesting that the models with the two bars
have a tendency to better reproduce the observed line-of-sight velocity of the
3-kpc arm (see Table~\ref{tab:vlosl0}). 
There is a big uncertainty in the model parameters, 
including the angular position of the bar in Case 2. Increasing the
angle from $20\gr$ to $40\gr$ makes the line-of-sight velocities in
the region of the 3-kpc arm to increase in absolute value,   
i.e. increases the rectangular shape of the orbits 
in the ($l$,$v$) diagram, while it decreases the size of the necklace-shaped 
feature corresponding to the GMR. In the three selected cases, we can see that 
several models can reach the observed line-of-sight velocities, and even exceed 
them, whereas Dehnen00 and GF10 clearly do not fit the observations. 

From Figs.~\ref{fig:COinvman}, \ref{fig:COinvman_2bars} and \ref{fig:COinvman_2bars_phi0} we observe that Fux01 in all three cases and MR09 
in cases 2 and 3 have velocities that highly exceed the 
terminal velocity curve and, therefore, they do not satisfy one of the main 
constraints of the ($l$,$v$)-diagram. This fact can be due to two reasons. 
First, these models have strong bars, i.e. the bar perturbation is quite 
noticeable in the outer parts. The second reason is morphological and
related to the previous one. The orbits trapped in the inner ring resemble 
the cuspy $x_1$ orbits of the bar, responsible for the maximum in the terminal 
velocity curves \citep{bin91}. So the inner branches of the orbits are in 
the immediate vicinity of the orbits of the bar, making an increase of the 
velocity at low longitudes. 

All the above discussion suggests that at least some of the model 
parameters may not be appropriate for the MW, and most of them may not be 
optimum. Indeed, they were selected in the papers described in the Appendix 
so as to give optimum results for the COBE/DIRBE bar only, and not for the two
bars together. We have nevertheless followed them here because they
are the standard reference models in the field. In a future paper we
will search for the parameter values which are optimum for the
COBE/DIRBE and Long bar system.

The spatial locus of the GMR is not well established. \citet{cle88} claim
that the peak emission of the GMR is at $\sim 5.5\,kpc$, while \citet{bin98},
\citet{dam01} and \citet{rat09} suggest that it is located around 
$\frac{1}{2}R_0$. \citet{cle88} believe that the GMR is an almost circular 
ring with a pitch angle of about $4^{\circ}$, while other authors 
\citep{fux99,dam01,rod08} favour the possibility that the molecular ring is 
composed of several inner spiral arm segments rooted in the central
bar. Kinematically,  
though, it is somewhat better constrained. It defines an over-density in 
the ($l$,$v$) diagram, crossing the plane from $\sim 100\,km\,s^{-1}$ and 
$l\sim 30^{\circ}$ to $\sim -100\,km\,s^{-1}$ and $l\sim -40^{\circ}$, passing 
through the origin and having a typical necklace shape. Morphologically, the 
outer branches of the manifolds in most of the cases have a size appropriate 
to the GMR, although Fux01, MR09 and Dehnen00 have a major diameter closer
to the Sun's position than to halfway from the Galactic Center. The shape 
varies from one model to another, the models of Dehnen00, PMM04 and GF10 
(with only one bar or with two bars) having the shape nearest to circular. 
Kinematically, not all models fit the enhancements related to the GMR. The 
manifold branches plotted in red in Figs.~\ref{fig:COinvman_2bars_phi0} and
\ref{fig:COinvman_2bars} (Cases 2 and 3) fit well inside the dashed lines in 
all models, while the dark blue branches lie inside the dashed lines only for 
models GF10 and PMM04. Fux01, MR09 and Dehnen00 the blue line makes a loop 
that resembles the over-density found between $l=30\gr-40\gr$ (see Fig.~5 of 
\citet{rat09}, where they plot the ($l$,$v$) diagram given by the $^{13}CO$ 
emission of the molecular clouds and clumps of the Boston Galactic Ring Survey).
Note, in addition, that the angular separation between the bars does 
not affect the global shape of the outer manifolds, either morphologically or 
in the ($l$,$v$) diagram.

\section{Obtaining the outer spiral arms}
\label{sec:outer}
As mentioned in the Introduction, one of the main features of the Milky Way 
is the presence of two stellar massive spiral arms \citep{ben05}. In this 
section, we discuss the COBE/DIRBE bar models focusing on the
parameters necessary in order for the manifolds to reproduce a
two-armed spiral shape. 

In Sect~\ref{sec:motion} and Paper II we showed that in a barred galaxy
potential, the two parameters that influence most the shape of the 
invariant manifolds in the outer parts are the bar mass (or amplitude of the 
m=2 component for quadrupole bars) and its pattern speed. In 
subsequent papers, we analysed the effect of the variation of both parameters 
and showed that the resultant gross global morphology is not model
dependent, and that, for all models,  
faster and more massive bars produce more open spiral arms, while slower 
and less massive bars produce symmetric outer rings (see Fig.~\ref{fig:2Dstudy}
and paper III).

In the previous section, we computed the invariant manifolds for five
selected MW models with the default parameters chosen by the authors. In 
all cases, the global morphology is that of an $rR_1$ ring whose extension 
does not reach the solar radius. As seen in Fig.~\ref{fig:2Dstudy}, if we 
want to reproduce the outer spiral arms of the Galaxy using orbits confined 
by the manifolds, we should significantly increase the bar mass or strength, 
and the pattern speed of each model. In Fig.~8 of Paper III, we plot the 
different models considered as a function of the strength parameter 
$Q_{t,L_1}$. We observed that the different morphologies are grouped as a 
function of $Q_{t,L_1}$, so if the bar model has $Q_{t,L_1}<0.1$ we have 
$R_1$ outer rings, while for $0.1<Q_{t,L_1}<0.2$ the model gives $R_1^{\prime}$ 
pseudo-rings, and for $Q_{t,L_1}>0.2$ it gives spirals and the other type 
of rings ($R_2$ and $R_1R_2$). So we need to use model parameters that 
give a value of $Q_{t,L_1}$ of around $0.2$, or larger. Note that this value is
significantly larger than the values given in Table~\ref{tab:forces}. We will 
see if this is a plausible option for the MW. 

In the case of quadrupole bars, the minimum values of $\alpha$ and
$\Omega_b$ necessary to obtain tightly wound spiral arms are $0.05$ and 
$50\,\ksp$, respectively (bottom panel of Fig.~\ref{fig:2Dstudy}). A model 
with two open spiral arms that crosses the solar radius can be obtained 
with $\alpha\ge 0.1$ and $\Omega_b\ge 40\,\ksp$. Even though the pattern 
speed falls within the range found by simulations \citep{eng99}, the bar 
strength is too high. Remember that $\alpha$ is the radial force along the 
bar semi-major axis and at the solar radius. It means that with a quadrupole 
model and a flat rotation curve, we need a bar whose strength is about 10\% 
of that of the disc at the solar radius. Besides, the $Q_b$ parameter of 
this model is $2.08$ far outside the observed range \citep{blo04}. If we 
vary the shape of the rotation curve and we make it slightly decreasing in 
the outer parts ($\beta=-0.2$), we can obtain two open spiral arms with 
$\alpha\ge 0.075$ and $\Omega_b\ge 57\,\ksp$. The radial force of the bar 
in the solar neighbourhood is still quite high but it gives $Q_b=0.63$, 
implying a strong bar, but within the observed range \citep{blo04}. This 
might indicate that if we want to reproduce the spiral arms using manifolds, 
it seems necessary to have a decreasing rotation curve in the outer parts. 
A recent study conducted by \citet{xue08} analysed more than $2400$ stars 
from the SDSS survey and concluded that it is indeed possible that the rotation 
curve of the MW decreases in the outer parts, by a factor within the 
range $\beta=(-0.2,-0.1)$, compatible with the suggested model. Other  
studies \citep{bra93} suggest that the rotation curve of the Milky Way could 
be slightly increasing in the outer parts, meaning that we should need 
stronger bars in order to obtain open spiral arms with manifolds.

In the case of Composite bars, the force decreases very abruptly 
in the outer parts of the disc, see Fig.~\ref{fig:forces}. In order to obtain 
values of $Q_{t,L_1}$ of the order of $0.1$ or $0.2$ we have to increase the 
pattern speed up to at least $\Omega_b\sim 75\,\ksp$ for a bar mass of 
$1.4 \times 10^{10}\,M_{\odot}$. If, on the other hand, we decrease the bar
axial ratio, also related to the bar mas, to $b/a=0.25$, we can obtain values
of $Q_{t,L_1}$ of the order of $0.1$ with a bar mass of 
$\sim 2.0 \times 10^{10}\,M_{\odot}$ and a pattern speed of 
$\Omega_b\sim 70\,\ksp$. Nevertheless, the value of $Q_{t,L_1}$ is still too low.
The case of Ferrers bars is analogue to the Composite bar. 

The main conclusion here is that it seems that in any case the COBE/DIRBE bar is
not strong enough to make two open spiral arms that could reach the outer
parts of the disc. Even though the introduction of the Long bar makes
the inner branches of the manifolds more elongated
(Sect.~\ref{sec:morphology}), it does not change the previous  
conclusions. Models with an extra component to the potential, for 
example a spiral forcing, should be explored in the future as a possible 
option to induce spiral arms through manifolds.

\section{Summary and Conclusions}
\label{sec:conc}

In this paper, we applied the invariant manifolds theory for the first 
time to the Milky Way. We first presented a thorough discussion about
whether the Long bar and the COBE/DIRBE bar are, respectively, the
primary and the secondary bar of a double bar system in our Galaxy. We
dismissed this alternative, because their properties are in strong
disagreement with the properties of double bars in external galaxies
and with those in simulations. We then
considered the alternative that the COBE/DIRBE 
bar and the Long bar are simply {\it parts of the same bar}, the former
being the boxy/peanut bulge and the latter the thin outer parts of the
bar. We compared the morphology thus obtained with that of external barred
galaxies, with that obtained from orbital structure and with
simulations, and found very good agreement in all cases. We also
proposed some reasonable possibilities to explain the fact that
observations indicate a misalignment between the major axes of the COBE/DIRBE 
bar and of the Long bar.

We then selected five characteristic models in the literature that include
the COBE/DIRBE bar, and, to account for recent observations and the 
above mentioned possibility, we have also
considered the case with an additional Long bar. We analyse the models, 
first in terms of forces, and then we compute the manifolds in three cases,
namely when only the COBE/DIRBE bar is modelled (Case 1), when the two bars 
are aligned at $30\gr$ from the Galactic centre - Sun line, as observed 
in a large amount of external galaxies (Case 2), and 
when the two bars have an angular separation of $20\gr$ according to
the observations (Case 3). Note, however, that in the models discussed here,
both bars rotate at the same pattern speed, which makes Case 3 dynamically 
unstable. 

Regarding the questions formulated in the Introduction, we confirm that 
the observed features of the inner parts of the Galaxy, more specifically 
the 3-kpc arms and the GMR, can be plausibly interpreted using manifolds.
The morphological and kinematic analysis shows that the inner and outer rings
described by the manifolds can well represent the 3-kpc arms and the GMR of
the inner parts of the Galaxy, respectively. Not all models, however, are 
suitable to describe both features at the same time and in both ways. 
In general, the GMR is well
reproduced by all models in all three cases, but the 3-kpc arms depend
more on the characteristics of the potential.

We can tentatively conclude that the two bars of the Galaxy, that
is, the COBE/DIRBE and the Long bar, are necessary to reproduce the
observables and, in this case, among the five 
models, the ones with less strong bars, namely PMM04 \citep{pic04}, 
Dehnen00 \citep{deh00} and GF10 \citep{gar10} have a better fit. Another 
conclusion of this work is that the value of the angular separation between the 
bars does introduce changes both in the morphology of the rings and in the
($l$,$v$) diagram.  

We also analysed in detail 
what bar parameters would be necessary so that the 
manifolds have a spiral arms morphology concluding that, for quadrupole bars,
a stronger bar is necessary, while for the composite and the Ferrers bars, 
the force decreases too abruptly in the outer parts of the disc, and it is
difficult to obtain open spiral arms by only increasing the bar mass or its
pattern speed. In the case, though, where the global morphology would be one 
with two spiral arms, the $rR_1$ configuration would be lost. This suggests 
that the most probable solution in the manifold framework would be a more 
complex potential with one or two bars in the inner part and  
a spiral further out (in preparation).

In a future paper we will revisit the most successful of the models
considered here and vary their main parameters -- i.e. the 
mass and the pattern speed of the bar, and the 
angle of the bar major axis with the Galactic center - Sun line --
searching for the values that give the best fits to the observations.

\section*{Acknowledgements}
We thank Albert Bosma for interesting discussions and encouragement,
Jean-Charles Lambert for his help with glnemo viewings and Peter Erwin 
for sharing with us his double bar data. We also thank J. Beckman, 
P. Garz\'on, P. L. Hammersley, M. L\'opez-Corredoira and T. J. Mahoney 
for very interesting discussions on the Long bar and on the basic ideas of
Sect.~\ref{sec:1or2bars}, following a seminar given by EA in IAC in 2006. This
work was supported by the MICINN (Spanish Ministry of Science and  
Innovation) - FEDER through grant AYA2009-14648-C02-01 and CONSOLIDER 
CSD2007-00050. TA acknowledges funding support from the European Research 
Council under ERC-StG grant GALACTICA-24027

\section*{Appendix}
\label{sec:app}
Here we describe the models we used throughout the paper, namely
models with a quadrupole bar \citep{fux01,deh00}, a composite bar 
\citep{pic04}, and a Ferrers bar \citep{fer77,mel09,gar10}. We also
give a brief description of the simulation used in Sect.~\ref{sec:1or2bars}.

\subsection*{The quadrupole bar model: Fux01 and Dehnen00}
The quadrupole model consists of the superposition of an axisymmetric 
component given by a simple power-law rotation curve and an $m=2$-type 
potential for the bar, as in \citet{deh00} and \citet{fux01}. We refer to
these models in the text as Dehnen00 and Fux01, respectively.

The potential corresponding to a power-law rotation curve is:

\begin{equation}\label{eq:rot}
\Phi_0(R)=v_0^2\left\{\begin{array}{ll}
(2\beta)^{-1}(R/R_0)^{2\beta}, & \beta\ne 0\\
\ln(R/R_0),                 &   \beta = 0,
\end{array}\right.
\end{equation}
where $R_0=8\,kpc$ denotes the galactocentric distance to the Sun and $v_0$ the
local circular speed. Since in this paper we compare the two Galaxy models, we 
will keep the original values of $R_0$ and $v_0$, namely $8\,kpc$ and
$220\,km\,s^{-1}$ for Dehnen00, and $8\,kpc$ and $200\,km\,s^{-1}$ 
for Fux01. The parameter $\beta$ is related to the shape of the rotation curve, 
with $\beta=0$ for a flat rotation curve, $\beta<0$ for a falling rotation curve 
and $\beta>0$ for a rising rotation curve. We use flat rotation curves
in this paper,
unless otherwise stated. In the case of falling rotation curves, we use a value
of $\beta=-0.2$.

The bar is described as the $m=2$ component of the Fourier decomposition
of the potential: $\Phi_b(R,\theta)=A(R)\cos(2\theta)$, where 

\begin{equation}
A(R)=A_b\left\{\begin{array}{ll}
\left(\frac{R}{R_b}\right)^3-2,  &  R\le R_b\\
-\left(\frac{R_b}{R}\right)^3,   &  R\ge R_b,
\end{array}\right.
\end{equation}
where $R_b$ and $A_b$ are the size and the amplitude of the bar. In both
models, $R_b=0.8r_{L_1}$ and $r_{L_1}=4.35\,kpc$. This Lagrangian radius 
together with the rotation curve given by Eq.~(\ref{eq:rot}), corresponds to 
a pattern speed of $\Omega_b=51\,\ksp$. 

The authors measure the strength parameter of the bar by the ratio of the 
forces due to the bar and to the axisymmetric power-law component at the Sun 
Galactocentric radius on the bar's semi-major axis. 

\begin{equation}
\alpha=3\frac{A_b}{v_0^2}\left( \frac{R_b}{R_0}\right)^3.
\end{equation} 
Note that this measure is dimensionless, it is directly related to the 
bar amplitude and it is related to Eq.~(\ref{eq:qr}) by $\alpha=Q_r(R_0)$.
When the Long bar is introduced in the potential, it is described also with an
$m=2$ component of the Fourier decomposition of the potential with 
$R_{bl}=0.92r_{L_1}$, $r_{L_1}=4.35\, kpc$ and $\alpha_l=0.6\alpha$.

\subsection*{The composite bar model: PMM04}
The composite bar is an analytical model designed to fit the density profile of 
the bar given by COBE/DIRBE. Again, it consists of the superposition 
of an axisymmetric component and a bar. The axisymmetric component is the 
result of the superposition of a bulge, a disc and a dark matter halo. The 
potential we used is based on the one considered by \citet{all91} to fit the 
axisymmetric component of the MW. The bulge and the disc are modelled 
using a Miyamoto-Nagai potential \citep{miy75}:

\begin{equation}
\Phi_{bl}(R,z)=-\frac{M_1}{\left (R^2+z^2+b_1^2\right)^{1/2}},
\end{equation}
where $(R,z)$ are the cylindrical coordinates, $M_1=1.4\times 10^{10}\,M_{\odot}$
is the bulge mass and $b_1=0.3873$ is the bulge scale-length.

\begin{equation}
\Phi_{d}(R,z)=-\frac{M_2}{\left (R^2+\left[ a_2+\left(z^2+b_2^2\right)^{1/2}\right]^2\right)^{1/2}},
\end{equation}
where $M_2=8.56\times 10^{10}\,M_{\odot}$ is the disc mass, $a_2=5.3178$ and 
$b_2=0.25$ are the radial and vertical scale-lengths, respectively.
The dark matter halo is described using a spherical potential:

\begin{eqnarray}
\lefteqn{\Phi_h(r)=-\frac{M(r)}{r}-} \nonumber \\
& & \frac{M_3}{1.02a_3}\left[ -\frac{1.02}{1+\left(\frac{r}{a_3}\right)^{1.02}}+\ln\left(1+\left(\frac{r}{a_3}\right)^{1.02}\right) \right]_r^{100},
\end{eqnarray}
where the halo radius is $100\,kpc$, $M_3=10.7\times 10^{10}\,M_{\odot}$, which 
makes the total mass of the halo $M(100\,kpc)=8.002\times 10^{11}M_{\odot}$, and 
$a_3=12$ is its scale-length.

The parameters are chosen so that the total mass of the axisymmetric component
is $9\times 10^{11}M_{\odot}$, the rotation curve flattens at approximately 
$200\,km\,s^{-1}$, setting the galactocentric distance to the Sun 
$R_0=8.5\,kpc$ and the circular velocity at the Sun's position to 
$v_0=220\,km\,s^{-1}$.

The bar component is taken from \citet{pic04}. The density distribution is 
obtained to match the observations from COBE/DIRBE. The bar is the result of 
the superposition of prolate spheroids with density:

\begin{equation}
\rho_b(R_s)=\rho_0\left\{\begin{array}{ll}
sech^2(R_s) & R_s\le R_{end_s}\\
sech^2(R_s)e^{-\left( \frac{(R_s-R_{end_S})^2}{h_{end_S}^2}\right )} & R_s\ge R_{end_s}
\end{array}\right.,
\end{equation}
where $R_s=\left(\frac{x^2}{a_x^2}+\frac{y^2+z^2}{a_y^2} \right)^{1/2}$. The 
parameters $a_x$ and $a_y$ are the scale-lengths of the bar and are fixed to
$1.7\,kpc$ and $0.54\,kpc$, respectively. The constants $R_{end_s}$ and $h_{end_s}$
are defined as $a_{bar}/a_x$ and $h_{end}/a_x$, respectively. The authors
fix $a_{bar}=3.13\,kpc$ and $h_{end}=0.46\,kpc$, as the values of the length
of the bar and its scale-height. The bar is divided in three regions to better
describe the density. The first two have a fall of $sech^2$ and the third
has a Gaussian fall that starts where the bar ends. This implies a steep but 
smooth decrease in the density in the outer parts. 
The bar mass is fixed to $M_b=10^{10}\,M_{\odot}$ and $\Omega_b=60\,\ksp$.
The model parameters of the Long bar, when included in the potential, are 
$a=4.5\,kpc$, $b/a=0.24$ and $M_{bl}=0.6M_b=6\times 10^9\,M_{\odot}$.

\subsection*{The Ferrers bar model: MR09 and GF10}
The third bar model we use is a Ferrers bar as used by \citet{mel09,gar10}.
The potential is described by the superposition of an axisymmetric plus a 
bar-like component. The axisymmetric component considered by the two
pairs of authors is
slightly different even though both have a flat rotation curve.

In \citet{mel09}, the axisymmetric part of the potential has two components:
a bulge and a halo. The bulge potential is a Plummer sphere \citep{bin08}. 
The bulge mass is fixed to $1.22\times 10^{10}\times M_{\odot}$ and
its scale-length to $0.31\,kpc$. The halo is described by its rotation curve
as

\begin{equation}
v^2(r)=v^2_{max}\frac{r^2}{r^2+r_c^2},
\end{equation}
where $v_{max}=251.6\,km\,s^{-1}$ is the asymptotic maximum of the halo rotation
curve and $r_c=8.\,kpc$ is its core radius.

In \citet{gar10}, the axisymmetric part of the potential has three components:
a bulge, a disc and a dark halo. The bulge is a superposition of two 
Plummer spheres, with masses $3\times 10^{9}\,M_{\odot}$ and 
$1.6\times 10^{10}\,M_{\odot}$, respectively, and scale-lengths $2.7\,kpc$ 
and $0.42\,kpc$, respectively. The halo is described by an axisymmetric 
logarithmic potential with asymptotic velocity fixed to $220\,km\,s^{-1}$ and 
core radius $8.5\,kpc$. Finally the disc results of the superposition of 
three Miyamoto-Nagai discs \citep{miy75} with masses: 
$7.704\times 10^{10}\,M_{\odot}$, $-6.848\times 10^{10}\,M_{\odot}$ and 
$2.675\times 10^{10}\,M_{\odot}$, respectively, and radial scale-lengths: 
$5.81\,kpc$, $17.43\,kpc$ and $34.84\,kpc$, respectively. The vertical 
scale-length is the same for the three discs and it is fixed to $0.3\,kpc$.

As mentioned above, in both models, the bar component is described by a 
Ferrers ellipsoid \citep{fer77}, whose density distribution is described by 
the expression

\begin{equation}
\rho=\left\{ \begin{array}{ll}
\rho_0(1-m^2)^n, & m\le 1\\
0,               & m> 1,
\end{array}
\right.
\end{equation}
where $m^2=x^2/a^2+y^2/b^2$. The parameter $n$ measures the degree of 
concentration of the bar and $\rho_0$ measures its central concentration. 
It is related to the bar mass via the expression:

\begin{equation}
M_b=2^{2n+3}\pi ab^2\rho_0 \Gamma(n+1)\Gamma(n+2)/\Gamma(2n+4).
\end{equation}

In \citet{mel09}, the authors fix $n=1$, $a=3.82\,kpc$, $b=1.2\,kpc$ and 
$M_b=1.82\times 10^{10}\,M_{\odot}$. The default value for the pattern speed is 
$\Omega_b=53\,\ksp$. In \citet{gar10}, the authors fix $n=2$, $a=3.5\,kpc$, 
$b=1.4\,kpc$ and $M_b=10^{10}\,M_{\odot}$, and the default value
for the pattern speed is $\Omega_b=55.9\,\ksp$.
The model parameters for the Long bar, when included in the potential, are
$a=4.5\,kpc$, $b/a=0.15\,kpc$ and $M_b=1.1\times 10^{10}\,M_{\odot}$ for MR09
and $a=3.9\,kpc$, $b/a=0.15\,kpc$ and $M_{bl}\sim 0.6M_b=6\times 10^{9}\,M_{\odot}$
for GF10.

\subsection*{The simulation in Sect.~\ref{sec:1or2bars}}
The simulation discussed in Sect.~\ref{sec:1or2bars} is part of the
library of $N$-body simulations run by EA, but was not made
specifically to model the Milky Way, and is only intended for
illustration purposes. In the initial conditions the disc has an
exponential horizontal profile and a $sech^2$ vertical one. The halo
is described by equation (2.2) of \cite{Hernquist.93} and the bulge by
a Hernquist sphere \citep{Hernquist.90}. It was run   
using the public version of the gyrfalcon code \citep{Dehnen2000:falcON,
  Dehnen02}. The resolution for Fig~\ref{fig:1or2bars} was enhanced
using the technique described by \cite{Athanassoula05} and made use
of the glnemo2 display software, written by J.-C. Lambert. The same
software was also used to create the short movies. 

\bibliography{manifGal}

\label{lastpage}

\end{document}